\documentclass{article}
\usepackage{graphicx,amssymb,amsmath,times}
\usepackage[backref=true,pagebackref=true]{hyperref}

\newcommand{\mcm}[3]{\newcommand{#1}[#2]{{\ensuremath{#3}}}}

\usepackage{latexsym}
\usepackage{amssymb}

\mcm{\blank}{0}{(\emptybk)} \mcm{\dashbk}{0}{\mbox{---}}
\mcm{\emptybk}{0}{\:\:} \mcm{\hyph}{0}{\mbox{-}}

\mcm{\diagspace}{0}{\mbox{\hspace{2em}}}

\mcm{\cat}{1}{\mc{#1}} \mcm{\fcat}{1}{\mb{#1}}
\mcm{\mc}{1}{\mathcal{#1}} \mcm{\mr}{1}{\mathrm{#1}}
\mcm{\mi}{1}{\mathit{#1}} \mcm{\mb}{1}{\mathbf{#1}}
\mcm{\scat}{1}{\Bbb{#1}} \mcm{\twid}{1}{\widetilde{#1}}

\mcm{\elt}{0}{\in} \mcm{\sub}{0}{\,\subseteq\,}
\mcm{\such}{0}{\:|\:} \mcm{\without}{0}{\setminus}

\mcm{\atsr}{0}{\Box} \mcm{\eqv}{0}{\,\simeq\,}
\mcm{\iso}{0}{\,\cong\,}
\mcm{\of}{0}{\raisebox{0.2mm}{\ensuremath{\scriptstyle\circ}}}

\mcm{\bdry}{0}{\partial}

\mcm{\Bee}{0}{\cat{B}} \mcm{\Beep}{0}{\cat{B'}}
\mcm{\Eee}{0}{\cat{E}} \mcm{\Eeep}{0}{\cat{E'}}

\mcm{\Ess}{0}{\cat{S}} \mcm{\Tee}{0}{\cat{T}}
\mcm{\Teep}{0}{\cat{T'}} \mcm{\Stee}{0}{\scat{T}}
\mcm{\Steep}{0}{\scat{T'}}

\mcm{\blbk}{0}{\blank^{\blob}}
\mcm{\blob}{0}{\scriptscriptstyle{\bullet}}
\mcm{\stbk}{0}{\blank^{*}} \mcm{\ubl}{0}{{}^{\blob}}
\mcm{\ust}{0}{{}^{*}}

\mcm{\Cartpr}{0}{\pr{\Eee}{T}} \mcm{\Cartprp}{0}{\pr{\Eeep}{T'}}
\mcm{\Mnd}{0}{\triple{T}{\eta}{\mu}}
\mcm{\Zeropr}{0}{\pr{\Set}{\id}}

\mcm{\dopset}{0}{\ftrcat{\Delta^{\op}}{\Set}}
\mcm{\tropset}{0}{\ftrcat{\fcat{TR}^{\op}}{\Set}}

\mcm{\cod}{0}{\mr{cod}} \mcm{\dom}{0}{\mr{dom}}
\mcm{\End}{0}{\mr{End}} \mcm{\Hom}{0}{\mr{Hom}}
\mcm{\ob}{0}{\mr{ob}\,} \mcm{\op}{0}{\mr{op}}

\mcm{\comp}{0}{\mi{comp}} \mcm{\id}{0}{\mi{id}}
\mcm{\ids}{0}{\mi{ids}} \mcm{\mult}{0}{\mi{mult}}
\mcm{\unit}{0}{\mi{unit}}

\mcm{\Ab}{0}{\fcat{Ab}} \mcm{\Alg}{0}{\fcat{Alg}}
\mcm{\Bim}{1}{\fcat{Bim}(#1)} \mcm{\Cat}{0}{\fcat{Cat}}
\mcm{\Cay}{0}{\fcat{Cay}} \mcm{\Cpn}{1}{\pr{\Set/S_{#1}}{T_{#1}}}
\mcm{\fc}{0}{\fcat{fc}} \mcm{\fm}{0}{\fcat{fm}}
\mcm{\Graph}{0}{\fcat{Graph}} \mcm{\Gy}{0}{\fcat{Gy}}
\mcm{\Hpn}{1}{\pr{\Eee_{#1}}{P_{#1}}} \mcm{\Mon}{0}{\mb{Mon}}
\mcm{\Multicat}{0}{\fcat{Multicat}} \mcm{\One}{0}{\fcat{1}}
\mcm{\PD}{1}{\fcat{PD}_{#1}} \mcm{\Prof}{0}{\fcat{Prof}}
\mcm{\Set}{0}{\fcat{Set}} \mcm{\Span}{0}{\fcat{Span}}
\mcm{\Ssq}{0}{\fcat{Ssq}} \mcm{\Struc}{0}{\fcat{Struc}}
\mcm{\Sym}{0}{\fcat{Sym}} \mcm{\TR}{1}{\fcat{TR}(#1)}
\mcm{\Tr}{0}{\fcat{Tr}} \mcm{\Twocat}{0}{\fcat{2\hyph\Cat}}

\mcm{\integers}{0}{\mathbb{Z}}

\mcm{\range}{2}{#1,\,\ldots\,,#2}
\mcm{\bftuple}{2}{\tuplebts{\range{#1}{#2}}}
\mcm{\tuple}{3}{\tuplebts{\range{#1,#2}{#3}}}
\mcm{\rttuple}{1}{\tuplebts{\,\ldots\,,#1}}
\mcm{\abftuple}{2}{\atuplebts{\range{#1}{#2}}}
\mcm{\atuple}{3}{\atuplebts{\range{#1,#2}{#3}}}
\mcm{\arttuple}{1}{\atuplebts{\,\ldots\,,#1}}
\mcm{\sqbftuple}{2}{\obt\range{#1}{#2}\cbt}
\mcm{\pr}{2}{\tuplebts{#1,#2}}
\mcm{\triple}{3}{\tuplebts{#1,#2,#3}}

\mcm{\eend}{2}{#1[#2]} \mcm{\ehom}{3}{#1[#2,#3]}
\mcm{\ftrcat}{2}{[#1,#2]} \mcm{\homset}{3}{#1(#2,#3)}
\mcm{\multihom}{3}{#1(#2;#3)}
\mcm{\relhom}{5}{#1_{#2}(\range{#3}{#4};#5)}

\mcm{\go}{0}{\rTo} \mcm{\goby}{1}{\rTo^{#1}}
\mcm{\goesto}{0}{\,\longmapsto\,} \mcm{\goiso}{0}{\goby{\diso}}
\mcm{\monic}{0}{\rMonic} \mcm{\og}{0}{\lTo}
\mcm{\ogby}{1}{\lTo^{#1}}

\mcm{\gph}{2}{\spn{#1}{T #2}{#2}} \mcm{\graph}{4}{\spaan{#1}{T
#2}{#2}{#3}{#4}} \mcm{\oppair}{2}{\stackrel{\rTo^{#1}}{\lTo_{#2}}}
\mcm{\parpair}{2}{\stackrel{\rTo^{#1}}{\rTo_{#2}}}
\mcm{\spn}{3}{#2 \og #1 \go #3} \mcm{\spaan}{5}{#2 \ogby{#4} #1
\goby{#5} #3}

\mcm{\bktdvslob}{3}
    {\left(
    \begin{diagram}[height=1.5em]
    #1      \\
    \dTo>{\,#2} \\
    #3      \\
    \end{diagram}
    \right)}
\mcm{\slob}{3}{(#1 \goby{#2} #3)} \mcm{\vslob}{3}
    {\left.
    \begin{diagram}[height=1.5em]
    #1      \\
    \dTo>{\,#2} \\
    #3      \\
    \end{diagram}
    \right.}

\newenvironment{tree}
    {\begin{diagram}[height=1em,width=.75em,abut,noPS,tight]}
    {\end{diagram}}

\mcm{\enode}{0}{\circ}

\mcm{\nl}{1}{\stackrel{\textstyle #1}{\node}}
\mcm{\node}{0}{\bullet}

\mcm{\utree}{0}{\node}

\mcm{\diso}{0}{\sim}

\mcm{\vdiso}{0}{\wr}

\mcm{\nat}{0}{\mathbb{N}}

\mcm{\Onepr}{0}{\pr{\Graph}{\fc}}
\newlength{\nllwidth}
\newlength{\nllheight}
\newcommand{\stackbelow}[2]{%
\settowidth{\nllwidth}{\ensuremath{#1}\ensuremath{#2}}%
\settoheight{\nllheight}{\ensuremath{#2}}%
\addtolength{\nllheight}{.3ex}%
\mbox{%
\ensuremath{#1}%
\hspace{-.5\nllwidth}%
\raisebox{-1\nllheight}{\ensuremath{#2}}}}
\mcm{\nlal}{2}{\stackbelow{\nl{#1}}{#2}}
\mcm{\nll}{1}{\stackbelow{\node}{#1}} \mcm{\wun}{0}{\fcat{1}}
\mcm{\atuplebts}{1}{\langle #1 \rangle} \mcm{\tuplebts}{1}{(#1)}
\mcm{\bo}{0}{(} \mcm{\bc}{0}{)}
\mcm{\UBilax}{0}{\fcat{UBicat}_\mr{lax}}
\mcm{\UBiwk}{0}{\fcat{UBicat}_\mr{wk}}
\mcm{\UBistr}{0}{\fcat{UBicat}_\mr{str}}
\mcm{\Bilax}{0}{\fcat{Bicat}_\mr{lax}}
\mcm{\Biwk}{0}{\fcat{Bicat}_\mr{wk}}
\mcm{\Bistr}{0}{\fcat{Bicat}_\mr{str}} \mcm{\rotsub}{0}{\cup
\raisebox{0.1em}{$\scriptstyle{|}$}} \mcm{\pd}{0}{\fcat{pd}}
\mcm{\rep}{1}{\widehat{#1}} \mcm{\ovln}{1}{\overline{#1}}
\mcm{\Gph}{0}{\fcat{Gph}} \mcm{\tr}{0}{\fcat{tr}}

\mcm{\ladj}{0}{\,\dashv\,} \mcm{\zeropd}{0}{\node}
    {\end{diagram}}
\mcm{\END}{0}{\fcat{End}} \mcm{\HOM}{0}{\fcat{Hom}}

\newlength{\gwidth} 
\newlength{\gvert}  
\newlength{\gdrop}  
\newlength{\gbaredrop}  
\newlength{\goffset}    
\newlength{\gtemp}  

\newcommand{\present}[1]{%
\makebox[1\gwidth]{%
\rule[-1\gdrop]{0ex}{1\gvert}%
\raisebox{-1\gbaredrop}{#1}}}

\newcommand{\presentl}[1]{%
\makebox[1\gwidth][l]{%
\rule[-1\gdrop]{0ex}{1\gvert}%
\raisebox{-1\gbaredrop}{#1}}}

\newcommand{\presentr}[1]{%
\makebox[1\gwidth][r]{%
\rule[-1\gdrop]{0ex}{1\gvert}%
\raisebox{-1\gbaredrop}{#1}}}

\newcommand{\ginitdims}[2]{
\setlength{\unitlength}{1em}
\setlength{\goffset}{.25\unitlength}
\setlength{\gwidth}{#1\unitlength}
\setlength{\gvert}{#2\unitlength}
\setlength{\gdrop}{.5\gvert}
\addtolength{\gdrop}{-1\goffset}
\setlength{\gbaredrop}{1\gdrop}
\addtolength{\gvert}{.6\unitlength}
\addtolength{\gdrop}{.3\unitlength}}    

\newcommand{\cinitdims}[2]{
\setlength{\unitlength}{1em}
\setlength{\goffset}{.35\unitlength}
\setlength{\gwidth}{#1\unitlength}
\setlength{\gvert}{#2\unitlength}
\setlength{\gdrop}{.5\gvert}
\addtolength{\gdrop}{-1\goffset}
\setlength{\gbaredrop}{1\gdrop}
\addtolength{\gvert}{.6\unitlength}
\addtolength{\gdrop}{.3\unitlength}}    

\newcommand{\gsinitdims}[2]{
\setlength{\unitlength}{0.5em}
\setlength{\goffset}{.25\unitlength}
\setlength{\gwidth}{#1\unitlength}
\setlength{\gvert}{#2\unitlength}
\setlength{\gdrop}{.5\gvert}
\addtolength{\gdrop}{-1\goffset}
\setlength{\gbaredrop}{1\gdrop}
\addtolength{\gvert}{.6\unitlength}
\addtolength{\gdrop}{.3\unitlength}}    

\newcommand{\sidespic}[1]{%
\settowidth{\gtemp}{\ensuremath{#1}}%
\addtolength{\gwidth}{1\gtemp}}

\newcommand{\abovepic}[1]{%
\settoheight{\gtemp}{\ensuremath{#1}}%
\addtolength{\gvert}{1\gtemp}%
\settodepth{\gtemp}{\ensuremath{#1}}%
\addtolength{\gvert}{1\gtemp}}

\newcommand{\belowpic}[1]{%
\settoheight{\gtemp}{\ensuremath{#1}}%
\addtolength{\gvert}{1\gtemp}%
\addtolength{\gdrop}{1\gtemp}%
\settodepth{\gtemp}{\ensuremath{#1}}%
\addtolength{\gvert}{1\gtemp}%
\addtolength{\gdrop}{1\gtemp}}

\newcommand{\cell}[4]{\put(#1,#2){\makebox(0,0)[#3]{\ensuremath{#4}}}}
\mcm{\zmark}{0}{\scriptstyle{\bullet}}

\newcommand{\pregfst}[1]{%
\begin{picture}(0.5,0.2)(-0.5,-0.2)%
\cell{-0.1}{-0.2}{tr}{#1}%
\cell{0}{0}{c}{\zmark}%
\end{picture}}

\mcm{\gfst}{1}{%
\ginitdims{0.5}{0.4}%
\sidespic{#1}%
\belowpic{#1}%
\presentr{\pregfst{#1}}}

\newcommand{\preglst}[1]{%
\begin{picture}(0.5,0.2)(0,-0.2)%
\cell{0.1}{-0.2}{tl}{#1}%
\cell{0.05}{0}{c}{\zmark}%
\end{picture}}

\mcm{\glst}{1}{%
\ginitdims{.5}{.4}%
\sidespic{#1}%
\belowpic{#1}%
\presentl{\preglst{#1}}}

\newcommand{\preglft}[1]{%
\begin{picture}(0,0.2)(0,-0.2)%
\cell{-0.1}{-0.2}{tr}{#1}%
\cell{0.05}{0}{c}{\zmark}%
\end{picture}}

\mcm{\glft}{1}{%
\ginitdims{0}{.4}%
\belowpic{#1}%
\present{\preglft{#1}}}

\newcommand{\pregrgt}[1]{%
\begin{picture}(0,0.2)(0,-0.2)%
\cell{0.1}{-0.2}{tl}{#1}%
\cell{0.05}{0}{c}{\zmark}%
\end{picture}}

\mcm{\grgt}{1}{%
\ginitdims{0}{.4}%
\belowpic{#1}%
\present{\pregrgt{#1}}}

\newcommand{\pregblw}[1]{%
\begin{picture}(0,0.3)(0,-0.3)
\cell{0}{-0.3}{t}{#1}%
\cell{0.05}{0}{c}{\zmark}%
\end{picture}}

\mcm{\gblw}{1}{%
\ginitdims{0}{.6}%
\belowpic{#1}%
\present{\pregblw{#1}}}

\newcommand{\pregfbw}[1]{%
\begin{picture}(0,0.65)(0,-0.65)
\cell{0}{-0.65}{t}{#1}%
\cell{0.05}{0}{c}{\zmark}%
\end{picture}}

\mcm{\gfbw}{1}{%
\ginitdims{0}{1.3}%
\belowpic{#1}%
\present{\pregfbw{#1}}}

\newcommand{\pregzero}[1]{%
\begin{picture}(0.8,0.4)(-0.4,-0.4)
\cell{0}{-0.4}{t}{#1}%
\cell{0}{0}{c}{\zmark}%
\end{picture}}

\mcm{\gzero}{1}{%
\ginitdims{0.8}{.6}%
\belowpic{#1}%
\sidespic{#1}%
\present{\pregzero{#1}}}

\newcommand{\pregone}[1]{%
\begin{picture}(5,0.4)(0,-0.2)%
\cell{2.5}{0.2}{b}{#1}%
\put(0,0){\vector(1,0){5}}%
\end{picture}}

\mcm{\gone}{1}{%
\ginitdims{5}{0.4}%
\abovepic{#1}%
\present{\pregone{#1}}}

\newcommand{\pregtwo}[3]{%
\begin{picture}(5,3.4)(0,-0.2)%
\cell{2.5}{3.2}{b}{#1}%
\cell{2.5}{-.2}{t}{#2}%
\cell{2.7}{1.5}{l}{#3}%
\qbezier(0,1.5)(2.5,4.5)(5,1.5)%
\qbezier(0,1.5)(2.5,-1.5)(5,1.5)%
\put(5,1.5){\vector(1,-1){0}}%
\put(5,1.5){\vector(1,1){0}}%
\put(2.5,2.5){\vector(0,-1){2}}%
\end{picture}}

\mcm{\gtwo}{3}{%
\ginitdims{5}{3.4}%
\abovepic{#1}%
\belowpic{#2}%
\present{\pregtwo{#1}{#2}{#3}}}

\newcommand{\pregthree}[5]{%
\begin{picture}(5,5.4)(0,-1.2)%
\cell{2.5}{4.2}{b}{#1}%
\cell{1.5}{1.7}{b}{#2}%
\cell{2.5}{-1.2}{t}{#3}%
\cell{2.7}{2.75}{l}{#4}%
\cell{2.7}{0.25}{l}{#5}%
\qbezier(0,1.5)(2.5,6.5)(5,1.5)%
\qbezier(0,1.5)(2.5,-3.5)(5,1.5)%
\put(0,1.5){\vector(1,0){5}}%
\put(2.5,3.5){\vector(0,-1){1.5}}%
\put(2.5,1){\vector(0,-1){1.5}}%
\put(5,1.5){\vector(1,-3){0}}%
\put(5,1.5){\vector(1,3){0}}%
\end{picture}}

\mcm{\gthree}{5}{%
\ginitdims{5}{5.4}%
\abovepic{#1}%
\belowpic{#3}%
\present{\pregthree{#1}{#2}{#3}{#4}{#5}}}

\newcommand{\pregfour}[7]{%
\begin{picture}(5,8.4)(0,-2.7)%
\cell{2.5}{5.7}{b}{#1}%
\cell{1.5}{2.8}{b}{#2}%
\cell{1.5}{0.2}{t}{#3}%
\cell{2.5}{-2.7}{t}{#4}%
\cell{2.7}{4.25}{l}{#5}%
\cell{2.7}{1.5}{l}{#6}%
\cell{2.7}{-1.25}{l}{#7}%
\qbezier(0,1.5)(2.5,9.5)(5,1.5)%
\qbezier(0,1.5)(2.5,4)(5,1.5)%
\qbezier(0,1.5)(2.5,-1)(5,1.5)%
\qbezier(0,1.5)(2.5,-6.5)(5,1.5)%
\put(2.5,5.25){\vector(0,-1){2}}%
\put(2.5,2.5){\vector(0,-1){2}}%
\put(2.5,-0.25){\vector(0,-1){2}}%
\put(5,1.5){\vector(1,-4){0}}%
\put(5,1.5){\vector(4,-3){0}}%
\put(5,1.5){\vector(4,3){0}}%
\put(5,1.5){\vector(1,4){0}}%
\end{picture}}

\mcm{\gfour}{7}{%
\ginitdims{5}{8.4}%
\abovepic{#1}%
\belowpic{#4}%
\present{\pregfour{#1}{#2}{#3}{#4}{#5}{#6}{#7}}}

\newcommand{\pregthreecell}[5]{%
\begin{picture}(8,5)(-4,-2.5)%
\cell{0}{2.5}{b}{#1}%
\cell{0}{-2.5}{t}{#2}%
\cell{-1.7}{0}{r}{#3}%
\cell{1.7}{0}{l}{#4}%
\cell{0}{0.2}{b}{#5}%
\qbezier(-4,0)(0,4.2)(4,0)%
\qbezier(-4,0)(0,-4.2)(4,0)%
\qbezier(-0.5,1.8)(-2.5,0)(-0.5,-1.8)%
\qbezier(0.5,1.8)(2.5,0)(0.5,-1.8)%
\put(-1,0){\vector(1,0){2}}%
\put(4,0){\vector(1,-1){0}}%
\put(4,0){\vector(1,1){0}}%
\put(-0.5,-1.8){\vector(1,-1){0}}%
\put(0.5,-1.8){\vector(-1,-1){0}}%
\end{picture}}

\mcm{\gthreecell}{5}{%
\ginitdims{8}{5}%
\abovepic{#1}%
\belowpic{#2}%
\present{\pregthreecell{#1}{#2}{#3}{#4}{#5}}}

\newcommand{\pregthreecellu}{%
\begin{picture}(5,3.4)(-0.5,-0.2)%
\qbezier(-.5,1.5)(2,4.5)(4.5,1.5)%
\qbezier(-.5,1.5)(2,-1.5)(4.5,1.5)%
\qbezier(1.5,2.7)(0.5,1.5)(1.5,0.3)%
\qbezier(2.5,2.7)(3.5,1.5)(2.5,0.3)%
\put(1.3,1.5){\vector(1,0){1.4}}%
\put(4.5,1.5){\vector(1,-1){0}}%
\put(4.5,1.5){\vector(1,1){0}}%
\put(1.5,0.3){\vector(2,-3){0}}%
\put(2.5,0.3){\vector(-2,-3){0}}%
\end{picture}}

\mcm{\gthreecellu}{0}{%
\ginitdims{5}{3.4}%
\present{\pregthreecellu}}

\newcommand{\pregtwocentre}[3]{%
\begin{picture}(5,3.4)(0,-0.2)%
\cell{2.5}{3.2}{b}{#1}%
\cell{2.5}{-.2}{t}{#2}%
\cell{2.5}{1.5}{c}{#3}%
\qbezier(0,1.5)(2.5,4.5)(5,1.5)%
\qbezier(0,1.5)(2.5,-1.5)(5,1.5)%
\put(5,1.5){\vector(1,-1){0}}%
\put(5,1.5){\vector(1,1){0}}%
\put(2.5,2.5){\vector(0,-1){2}}%
\end{picture}}

\mcm{\gtwocentre}{3}{%
\ginitdims{5}{3.4}%
\abovepic{#1}%
\belowpic{#2}%
\present{\pregtwocentre{#1}{#2}{#3}}}

\newcommand{\pregspecialone}[9]{%
\begin{picture}(8,8)(-4,-4)%
\cell{0}{3.9}{b}{#1}%
\cell{-2}{-0.2}{t}{#2}%
\cell{0}{-3.9}{t}{#3}%
\cell{-1.5}{1.1}{r}{#4}%
\cell{0.2}{1.5}{l}{#5}%
\cell{1.5}{1.1}{l}{#6}%
\cell{0.2}{-2}{l}{#7}%
\cell{-0.9}{2.3}{b}{#8}%
\cell{0.9}{2.3}{b}{#9}%
\qbezier(-4,0)(0,8)(4,0)%
\qbezier(-4,0)(0,-8)(4,0)%
\qbezier(-0.5,3.4)(-3.5,2)(-0.5,0.6)%
\qbezier(0.5,3.4)(3.5,2)(0.5,0.6)%
\put(-4,0){\vector(1,0){8}}%
\put(0,3.4){\vector(0,-1){2.8}}%
\put(0,-0.8){\vector(0,-1){2.4}}%
\put(-1.5,2.2){\vector(1,0){1.2}}%
\put(0.3,2.2){\vector(1,0){1.2}}%
\put(4,0){\vector(1,-2){0}}%
\put(4,0){\vector(1,2){0}}%
\put(-0.5,0.6){\vector(2,-1){0}}%
\put(0.5,0.6){\vector(-2,-1){0}}%
\end{picture}}

\mcm{\gspecialone}{9}{%
\ginitdims{8}{8}%
\abovepic{#1}%
\belowpic{#3}%
\present{\pregspecialone{#1}{#2}{#3}{#4}{#5}{#6}{#7}{#8}{#9}}}

\newcommand{\pregspecialtwo}{%
\begin{picture}(5,3.4)(0,-0.2)%
\qbezier(0,1.5)(2.5,4.5)(5,1.5)%
\qbezier(0,1.5)(2.5,-1.5)(5,1.5)%
\qbezier(1.7,2.5)(0,1.5)(1.7,0.5)%
\qbezier(3.3,2.5)(5,1.5)(3.3,0.5)%
\put(5,1.5){\vector(1,-1){0}}%
\put(5,1.5){\vector(1,1){0}}%
\put(1.7,0.5){\vector(3,-2){0}}%
\put(3.3,0.5){\vector(-3,-2){0}}%
\put(2.5,2.5){\vector(0,-1){2}}%
\put(1.2,1.5){\vector(1,0){1}}%
\put(2.8,1.5){\vector(1,0){1}}%
\end{picture}}

\mcm{\gspecialtwo}{0}{%
\ginitdims{5}{3.4}%
\present{\pregspecialtwo}}

\newcommand{\pregspecialthree}{%
\begin{picture}(5,5.4)(0,-1.2)%
\qbezier(0,1.5)(2.5,6.5)(5,1.5)%
\qbezier(0,1.5)(2.5,-3.5)(5,1.5)%
\qbezier(2,3.5)(1,2.75)(2,2)%
\qbezier(3,3.5)(4,2.75)(3,2)%
\qbezier(2,1)(1,0.25)(2,-0.5)%
\qbezier(3,1)(4,0.25)(3,-0.5)%
\put(0,1.5){\vector(1,0){5}}%
\put(1.5,2.75){\vector(1,0){2}}%
\put(1.5,0.25){\vector(1,0){2}}%
\put(5,1.5){\vector(1,-3){0}}%
\put(5,1.5){\vector(1,3){0}}%
\put(2,2){\vector(1,-1){0}}%
\put(3,2){\vector(-1,-1){0}}%
\put(2,-0.5){\vector(1,-1){0}}%
\put(3,-0.5){\vector(-1,-1){0}}%
\end{picture}}

\mcm{\gspecialthree}{0}{%
\ginitdims{5}{5.4}%
\present{\pregspecialthree}}

\newcommand{\pregonew}[1]{%
\begin{picture}(8,0.4)(0,-0.2)%
\cell{4}{0.2}{b}{#1}%
\put(0,0){\vector(1,0){8}}%
\end{picture}}

\mcm{\gonew}{1}{%
\ginitdims{8}{0.4}%
\abovepic{#1}%
\present{\pregonew{#1}}}

\mcm{\gzersu}{0}{%
\gsinitdims{0}{.6}%
\present{\pregblw{}}}

\mcm{\gonesu}{0}{%
\gsinitdims{5}{0.4}%
\present{\pregone{}}}

\mcm{\gtwosu}{0}{%
\gsinitdims{5}{3.4}%
\present{\pregtwo{}{}{}}}

\mcm{\gthreesu}{0}{%
\gsinitdims{5}{5.4}%
\present{\pregthree{}{}{}{}{}}}

\mcm{\gfoursu}{0}{%
\gsinitdims{5}{8.4}%
\present{\pregfour{}{}{}{}{}{}{}}}

\newcommand{\precone}[1]{%
\begin{picture}(4.2,0.4)(-0.3,-0.2)%
\cell{1.8}{0.2}{b}{#1}%
\put(0,0){\vector(1,0){3.6}}%
\end{picture}}

\mcm{\cone}{1}{%
\cinitdims{4.2}{0.4}%
\abovepic{#1}%
\present{\precone{#1}}}

\mcm{\gfstsu}{0}{%
\gsinitdims{0.5}{0.4}%
\presentr{\pregfst{}}}

\mcm{\glstsu}{0}{%
\gsinitdims{0.5}{0.4}%
\presentl{\preglst{}}}

\newcommand{\prectwodbl}[3]%
{\begin{picture}(4.2,3.4)(-0.1,-0.2)%
\cell{2}{3.2}{b}{#1}%
\cell{2}{-0.2}{t}{#2}%
\cell{2.3}{1.5}{l}{#3}%
\qbezier(0,2)(2,4)(4,2)%
\qbezier(0,1)(2,-1)(4,1)%
\put(4,2){\vector(1,-1){0}}%
\put(4,1){\vector(1,1){0}}%
\put(1.9,2.5){\line(0,-1){1.8}}%
\put(2.1,2.5){\line(0,-1){1.8}}%
\cell{2.01}{0.4}{b}{\vee}%
\end{picture}}

\mcm{\ctwodbl}{3}{%
\cinitdims{4.2}{3.4}%
\abovepic{#1}%
\belowpic{#2}%
\present{\prectwodbl{#1}{#2}{#3}}}

\newcommand{\precthreedbl}[5]{%
\begin{picture}(4.2,5.4)(-0.1,-0.2)%
\cell{2}{5.2}{b}{#1}%
\cell{1}{2.7}{b}{#2}%
\cell{2}{-.2}{t}{#3}%
\cell{2.3}{3.75}{l}{#4}%
\cell{2.3}{1.25}{l}{#5}%
\qbezier(0,3)(2,7)(4,3)%
\qbezier(0,2)(2,-2)(4,2)%
\put(0,2.5){\vector(1,0){4}}%
\put(1.9,4.5){\line(0,-1){1.3}}%
\put(2.1,4.5){\line(0,-1){1.3}}%
\cell{2.01}{2.9}{b}{\vee}%
\put(1.9,2){\line(0,-1){1.3}}%
\put(2.1,2){\line(0,-1){1.3}}%
\cell{2.01}{0.4}{b}{\vee}%
\put(4,3){\vector(1,-3){0}}%
\put(4,2){\vector(1,3){0}}%
\end{picture}}

\mcm{\cthreedbl}{5}{%
\cinitdims{4.2}{5.4}%
\abovepic{#1}%
\belowpic{#3}%
\present{\precthreedbl{#1}{#2}{#3}{#4}{#5}}}

\newcommand{\precthreecelltrp}[5]{%
\begin{picture}(8.2,5)(-4.1,-2.5)%
\cell{0}{2.5}{b}{#1}%
\cell{0}{-2.5}{t}{#2}%
\cell{-1.8}{0}{r}{#3}%
\cell{1.8}{0}{l}{#4}%
\cell{0}{0.3}{b}{#5}%
\qbezier(-4,0.5)(0,4)(4,0.5)%
\qbezier(-4,-0.5)(0,-4)(4,-0.5)%
\qbezier(-0.6,2)(-2.6,0)(-0.6,-2)%
\qbezier(-0.4,2)(-2.4,0)(-0.5,-1.9)%
\cell{-0.6}{-2}{b}{\lrcorner}%
\qbezier(0.4,2)(2.4,0)(0.5,-1.9)%
\qbezier(0.6,2)(2.6,0)(0.6,-2)%
\cell{0.65}{-2}{b}{\llcorner}%
\put(-1,0.15){\line(1,0){1.7}}%
\put(-1,0){\line(1,0){2}}%
\put(-1,-0.15){\line(1,0){1.7}}%
\cell{1.15}{0}{r}{>}%
\put(4,0.5){\vector(1,-1){0}}%
\put(4,-0.5){\vector(1,1){0}}%
\end{picture}}

\mcm{\cthreecelltrp}{5}{%
\cinitdims{8.2}{5}%
\abovepic{#1}%
\belowpic{#2}%
\present{\precthreecelltrp{#1}{#2}{#3}{#4}{#5}}}

\newcommand{\prectwo}[3]%
{\begin{picture}(4.2,3.4)(-0.1,-0.2)%
\cell{2}{3.2}{b}{#1}%
\cell{2}{-0.2}{t}{#2}%
\cell{2.2}{1.5}{l}{#3}%
\qbezier(0,2)(2,4)(4,2)%
\qbezier(0,1)(2,-1)(4,1)%
\put(4,2){\vector(1,-1){0}}%
\put(4,1){\vector(1,1){0}}%
\put(2,2.5){\vector(0,-1){2}}%
\end{picture}}

\mcm{\ctwo}{3}{%
\cinitdims{4.2}{3.4}%
\abovepic{#1}%
\belowpic{#2}%
\present{\prectwo{#1}{#2}{#3}}}

\newcommand{\precthree}[5]{%
\begin{picture}(4.2,5.4)(-0.1,-0.2)%
\cell{2}{5.2}{b}{#1}%
\cell{1}{2.7}{b}{#2}%
\cell{2}{-.2}{t}{#3}%
\cell{2.2}{3.75}{l}{#4}%
\cell{2.2}{1.25}{l}{#5}%
\qbezier(0,3)(2,7)(4,3)%
\qbezier(0,2)(2,-2)(4,2)%
\put(0,2.5){\vector(1,0){4}}%
\put(2,4.5){\vector(0,-1){1.5}}%
\put(2,2){\vector(0,-1){1.5}}%
\put(4,3){\vector(1,-3){0}}%
\put(4,2){\vector(1,3){0}}%
\end{picture}}

\mcm{\cthree}{5}{%
\cinitdims{4.2}{5.4}%
\abovepic{#1}%
\belowpic{#3}%
\present{\precthree{#1}{#2}{#3}{#4}{#5}}}

\newcommand{\prectwoop}[3]%
{\begin{picture}(4.2,3.4)(-0.1,-0.2)%
\cell{2}{3.2}{b}{#1}%
\cell{2}{-0.2}{t}{#2}%
\cell{2.2}{1.5}{l}{#3}%
\qbezier(0,2)(2,4)(4,2)%
\qbezier(0,1)(2,-1)(4,1)%
\put(0,2){\vector(-1,-1){0}}%
\put(0,1){\vector(-1,1){0}}%
\put(2,2.5){\vector(0,-1){2}}%
\end{picture}}

\mcm{\ctwoop}{3}{%
\cinitdims{4.2}{3.4}%
\abovepic{#1}%
\belowpic{#2}%
\present{\prectwoop{#1}{#2}{#3}}}

\newcommand{\prectwopar}[4]{%
\begin{picture}(4.2,3.4)(-0.1,-0.2)%
\cell{2}{3.2}{b}{#1}%
\cell{2}{-0.2}{t}{#2}%
\cell{1.6}{1.5}{r}{#3}%
\cell{2.4}{1.5}{l}{#4}%
\qbezier(0,2)(2,4)(4,2)%
\qbezier(0,1)(2,-1)(4,1)%
\put(4,2){\vector(1,-1){0}}%
\put(4,1){\vector(1,1){0}}%
\put(1.8,2.5){\vector(0,-1){2}}%
\put(2.2,2.5){\vector(0,-1){2}}%
\end{picture}}

\mcm{\ctwopar}{4}{%
\cinitdims{4.2}{3.4}%
\abovepic{#1}%
\belowpic{#2}%
\present{\prectwopar{#1}{#2}{#3}{#4}}}

\newcommand{\precthreein}[5]{%
\begin{picture}(4.2,5.4)(-0.1,-0.2)%
\cell{2}{5.2}{b}{#1}%
\cell{1}{2.7}{b}{#2}%
\cell{2}{-.2}{t}{#3}%
\cell{2.2}{3.75}{l}{#4}%
\cell{2.2}{1.25}{l}{#5}%
\qbezier(0,3)(2,7)(4,3)%
\qbezier(0,2)(2,-2)(4,2)%
\put(0,2.5){\vector(1,0){4}}%
\put(2,4.5){\vector(0,-1){1.5}}%
\put(2,0.5){\vector(0,1){1.5}}%
\put(4,3){\vector(1,-3){0}}%
\put(4,2){\vector(1,3){0}}%
\end{picture}}

\mcm{\cthreein}{5}{%
\cinitdims{4.2}{5.4}%
\abovepic{#1}%
\belowpic{#3}%
\present{\precthreein{#1}{#2}{#3}{#4}{#5}}}

\newcommand{\precthreecell}[5]{%
\begin{picture}(8.2,5)(-4.1,-2.5)%
\cell{0}{2.5}{b}{#1}%
\cell{0}{-2.5}{t}{#2}%
\cell{-1.7}{0}{r}{#3}%
\cell{1.7}{0}{l}{#4}%
\cell{0}{0.2}{b}{#5}%
\qbezier(-4,0.5)(0,4)(4,0.5)%
\qbezier(-4,-0.5)(0,-4)(4,-0.5)%
\qbezier(-0.5,2)(-2.5,0)(-0.5,-2)%
\qbezier(0.5,2)(2.5,0)(0.5,-2)%
\put(-1,0){\vector(1,0){2}}%
\put(4,0.5){\vector(1,-1){0}}%
\put(4,-0.5){\vector(1,1){0}}%
\put(-0.5,-2){\vector(1,-1){0}}%
\put(0.5,-2){\vector(-1,-1){0}}%
\end{picture}}

\mcm{\cthreecell}{5}{%
\cinitdims{8.2}{5}%
\abovepic{#1}%
\belowpic{#2}%
\present{\precthreecell{#1}{#2}{#3}{#4}{#5}}}

\newcommand{\precthreecellpar}[6]{%
\begin{picture}(8.2,5)(-4.1,-2.5)%
\cell{0}{2.5}{b}{#1}%
\cell{0}{-2.5}{t}{#2}%
\cell{-1.7}{0}{r}{#3}%
\cell{1.7}{0}{l}{#4}%
\cell{0}{0.4}{b}{#5}%
\cell{0}{-0.4}{t}{#6}%
\qbezier(-4,0.5)(0,4)(4,0.5)%
\qbezier(-4,-0.5)(0,-4)(4,-0.5)%
\qbezier(-0.5,2)(-2.5,0)(-0.5,-2)%
\qbezier(0.5,2)(2.5,0)(0.5,-2)%
\put(-1,0.2){\vector(1,0){2}}%
\put(-1,-0.2){\vector(1,0){2}}%
\put(4,0.5){\vector(1,-1){0}}%
\put(4,-0.5){\vector(1,1){0}}%
\put(-0.5,-2){\vector(1,-1){0}}%
\put(0.5,-2){\vector(-1,-1){0}}%
\end{picture}}

\mcm{\cthreecellpar}{6}{%
\cinitdims{8.2}{5}%
\abovepic{#1}%
\belowpic{#2}%
\present{\precthreecellpar{#1}{#2}{#3}{#4}{#5}{#6}}}

\newcommand{\prectwov}[5]{%
\begin{picture}(3.4,4.2)(0.8,0.9)%
\cell{2.5}{5.1}{b}{#1}%
\cell{2.5}{0.9}{t}{#2}%
\cell{0.8}{3}{r}{#3}%
\cell{4.2}{3}{l}{#4}%
\cell{2.5}{3.2}{b}{#5}%
\qbezier(2,5)(0,3)(2,1)%
\qbezier(3,5)(5,3)(3,1)%
\put(2,1){\vector(1,-1){0}}%
\put(3,1){\vector(-1,-1){0}}%
\put(1.5,3){\vector(1,0){2}}%
\end{picture}}

\mcm{\ctwov}{5}{%
\cinitdims{3.4}{4.2}%
\abovepic{#1}%
\belowpic{#2}%
\sidespic{#3}%
\sidespic{#4}%
\present{\prectwov{#1}{#2}{#3}{#4}{#5}}}

\newcommand{\precthreecellv}[7]{%
\begin{picture}(5,8.2)(0.5,-1.6)%
\cell{3}{6.6}{b}{#1}%
\cell{3}{-1.6}{t}{#2}%
\cell{0.5}{2.5}{r}{#3}%
\cell{5.5}{2.5}{l}{#4}%
\cell{3}{4.2}{b}{#5}%
\cell{3}{0.8}{t}{#6}%
\cell{3.2}{2.5}{l}{#7}%
\qbezier(3.5,6.5)(7,2.5)(3.5,-1.5)%
\qbezier(2.5,6.5)(-1,2.5)(2.5,-1.5)%
\put(2.5,-1.5){\vector(1,-1){0}}%
\put(3.5,-1.5){\vector(-1,-1){0}}%
\qbezier(1,3)(3,5)(5,3)%
\qbezier(1,2)(3,0)(5,2)%
\put(5,3){\vector(1,-1){0}}%
\put(5,2){\vector(1,1){0}}%
\put(3,3.5){\vector(0,-1){2}}%
\end{picture}}

\mcm{\cthreecellv}{7}{%
\cinitdims{5}{8.2}%
\abovepic{#1}%
\belowpic{#2}%
\sidespic{#3}%
\sidespic{#4}%
\present{\precthreecellv{#1}{#2}{#3}{#4}{#5}{#6}{#7}}}

\newcommand{\pretopez}[2]{%
\begin{picture}(2.6,2.3)(-1.3,-2.2)%
\cell{0}{-2.2}{t}{#1}%
\cell{0}{-1.2}{c}{#2}%
\qbezier(0,0)(-2,-2)(0,-2)%
\qbezier(0,0)(2,-2)(0,-2)%
\put(0,0){\vector(-1,1){0}}%
\end{picture}}

\mcm{\topez}{2}{%
\ginitdims{2.6}{2.3}%
\belowpic{#1}%
\present{\pretopez{#1}{#2}}}

\newcommand{\pretopea}[3]{%
\begin{picture}(4,1.9)(-2,-0,2)%
\cell{0}{1.7}{b}{#1}%
\cell{0}{-0.2}{t}{#2}%
\cell{0}{0.7}{c}{#3}%
\qbezier(-2,0)(0,3)(2,0)%
\put(-2,0){\vector(1,0){4}}%
\put(2,0){\vector(2,-3){0}}%
\end{picture}}

\mcm{\topea}{3}{%
\ginitdims{4}{1.9}%
\abovepic{#1}%
\belowpic{#2}%
\present{\pretopea{#1}{#2}{#3}}}

\newcommand{\pretopeb}[4]{%
\begin{picture}(4,2.2)(-2,-0.2)%
\cell{-1.1}{1}{br}{#1}%
\cell{1.1}{1}{bl}{#2}%
\cell{0}{-0.2}{t}{#3}%
\cell{0}{0.8}{c}{#4}%
\put(-2,0){\vector(1,1){2}}%
\put(0,2){\vector(1,-1){2}}%
\put(-2,0){\vector(1,0){4}}%
\end{picture}}

\mcm{\topeb}{4}{%
\ginitdims{4}{2.2}%
\belowpic{#3}%
\present{\pretopeb{#1}{#2}{#3}{#4}}}

\newcommand{\pretopec}[5]{%
\begin{picture}(4,2.2)(-2,-0.2)%
\cell{-1.8}{1}{br}{#1}%
\cell{0}{2.2}{b}{#2}%
\cell{1.8}{1}{bl}{#3}%
\cell{0}{-0.2}{t}{#4}%
\cell{0}{0.8}{c}{#5}%
\put(-2,0){\vector(1,2){1}}%
\put(-1,2){\vector(1,0){2}}%
\put(1,2){\vector(1,-2){1}}%
\put(-2,0){\vector(1,0){4}}%
\end{picture}}

\mcm{\topec}{5}{%
\ginitdims{4}{2.2}%
\sidespic{#1}%
\abovepic{#2}%
\sidespic{#3}%
\belowpic{#4}%
\present{\pretopec{#1}{#2}{#3}{#4}{#5}}}

\newcommand{\pretoped}[6]{%
\begin{picture}(4,2.5)(-2,-0.2)%
\cell{-2}{0.6}{br}{#1}%
\cell{-0.7}{2.2}{br}{#2}%
\cell{0.7}{2.2}{bl}{#3}%
\cell{2}{0.6}{bl}{#4}%
\cell{0}{-0.2}{t}{#5}%
\cell{0}{0.8}{c}{#6}%
\put(-2,0){\vector(1,3){0.5}}%
\put(-1.5,1.5){\vector(3,2){1.5}}%
\put(0,2.5){\vector(3,-2){1.5}}%
\put(1.5,1.5){\vector(1,-3){0.5}}%
\put(-2,0){\vector(1,0){4}}%
\end{picture}}

\mcm{\toped}{6}{%
\ginitdims{4}{2.5}%
\sidespic{#1}%
\abovepic{#2}%
\abovepic{#3}%
\sidespic{#4}%
\belowpic{#5}%
\present{\pretoped{#1}{#2}{#3}{#4}{#5}{#6}}}

\newcommand{\pretopeq}[5]{%
\begin{picture}(4,2.5)(-2,-0.2)%
\cell{-2}{0.6}{br}{#1}%
\cell{-1}{2.2}{br}{#2}%
\cell{2}{0.6}{bl}{#3}%
\cell{0}{-0.2}{t}{#4}%
\cell{0}{0.8}{c}{#5}%
\put(-2,0){\vector(1,3){0.5}}%
\put(-1.5,1.5){\vector(1,1){1}}%
\cell{0.9}{2.3}{c}{\ddots}
\put(1.5,1.5){\vector(1,-3){0.5}}%
\put(-2,0){\vector(1,0){4}}%
\end{picture}}

\mcm{\topeq}{5}{%
\ginitdims{4}{2.5}%
\sidespic{#1}%
\abovepic{#2}%
\sidespic{#3}%
\belowpic{#4}%
\present{\pretopeq{#1}{#2}{#3}{#4}{#5}}}

\newcommand{\pretopebase}[1]{%
\begin{picture}(4,0.4)(0,-0.2)%
\cell{2}{0.2}{b}{#1}%
\put(0,0){\vector(1,0){4}}%
\end{picture}}

\mcm{\topebase}{1}{%
\ginitdims{4}{0.4}%
\abovepic{#1}%
\present{\pretopebase{#1}}}

\newcommand{\pretopezs}[2]{%
\begin{picture}(2.6,2.3)(-1.3,-2.2)%
\cell{0}{-2.2}{t}{#1}%
\cell{0}{-1.2}{c}{#2}%
\qbezier(0,0)(-2,-2)(0,-2)%
\qbezier(0,0)(2,-2)(0,-2)%
\end{picture}}

\mcm{\topezs}{2}{%
\ginitdims{2.6}{2.3}%
\belowpic{#1}%
\present{\pretopezs{#1}{#2}}}

\newcommand{\pretopeas}[3]{%
\begin{picture}(4,1.9)(-2,-0,2)%
\cell{0}{1.7}{b}{#1}%
\cell{0}{-0.2}{t}{#2}%
\cell{0}{0.7}{c}{#3}%
\qbezier(-2,0)(0,3)(2,0)%
\put(-2,0){\line(1,0){4}}%
\end{picture}}

\mcm{\topeas}{3}{%
\ginitdims{4}{1.9}%
\abovepic{#1}%
\belowpic{#2}%
\present{\pretopeas{#1}{#2}{#3}}}

\newcommand{\pretopebs}[4]{%
\begin{picture}(4,2.2)(-2,-0.2)%
\cell{-1.1}{1}{br}{#1}%
\cell{1.1}{1}{bl}{#2}%
\cell{0}{-0.2}{t}{#3}%
\cell{0}{0.8}{c}{#4}%
\put(-2,0){\line(1,1){2}}%
\put(0,2){\line(1,-1){2}}%
\put(-2,0){\line(1,0){4}}%
\end{picture}}

\mcm{\topebs}{4}{%
\ginitdims{4}{2.2}%
\belowpic{#3}%
\present{\pretopebs{#1}{#2}{#3}{#4}}}

\newcommand{\pretopecs}[5]{%
\begin{picture}(4,2.2)(-2,-0.2)%
\cell{-1.8}{1}{br}{#1}%
\cell{0}{2.2}{b}{#2}%
\cell{1.8}{1}{bl}{#3}%
\cell{0}{-0.2}{t}{#4}%
\cell{0}{0.8}{c}{#5}%
\put(-2,0){\line(1,2){1}}%
\put(-1,2){\line(1,0){2}}%
\put(1,2){\line(1,-2){1}}%
\put(-2,0){\line(1,0){4}}%
\end{picture}}

\mcm{\topecs}{5}{%
\ginitdims{4}{2.2}%
\sidespic{#1}%
\abovepic{#2}%
\sidespic{#3}%
\belowpic{#4}%
\present{\pretopecs{#1}{#2}{#3}{#4}{#5}}}

\newcommand{\pretopeds}[6]{%
\begin{picture}(4,2.5)(-2,-0.2)%
\cell{-2}{0.6}{br}{#1}%
\cell{-0.7}{2.2}{br}{#2}%
\cell{0.7}{2.2}{bl}{#3}%
\cell{2}{0.6}{bl}{#4}%
\cell{0}{-0.2}{t}{#5}%
\cell{0}{0.8}{c}{#6}%
\put(-2,0){\line(1,3){0.5}}%
\put(-1.5,1.5){\line(3,2){1.5}}%
\put(0,2.5){\line(3,-2){1.5}}%
\put(1.5,1.5){\line(1,-3){0.5}}%
\put(-2,0){\line(1,0){4}}%
\end{picture}}

\mcm{\topeds}{6}{%
\ginitdims{4}{2.5}%
\sidespic{#1}%
\abovepic{#2}%
\abovepic{#3}%
\sidespic{#4}%
\belowpic{#5}%
\present{\pretopeds{#1}{#2}{#3}{#4}{#5}{#6}}}

\newcommand{\pretopeqs}[5]{%
\begin{picture}(4,2.5)(-2,-0.2)%
\cell{-2}{0.6}{br}{#1}%
\cell{-1}{2.2}{br}{#2}%
\cell{2}{0.6}{bl}{#3}%
\cell{0}{-0.2}{t}{#4}%
\cell{0}{0.8}{c}{#5}%
\put(-2,0){\line(1,3){0.5}}%
\put(-1.5,1.5){\line(1,1){1}}%
\cell{0.9}{2.3}{c}{\ddots}
\put(1.5,1.5){\line(1,-3){0.5}}%
\put(-2,0){\line(1,0){4}}%
\end{picture}}

\mcm{\topeqs}{5}{%
\ginitdims{4}{2.5}%
\sidespic{#1}%
\abovepic{#2}%
\sidespic{#3}%
\belowpic{#4}%
\present{\pretopeqs{#1}{#2}{#3}{#4}{#5}}}

\newcommand{\pretopebases}[1]{%
\begin{picture}(4,0.4)(0,-0.2)%
\cell{2}{0.2}{b}{#1}%
\put(0,0){\line(1,0){4}}%
\end{picture}}

\mcm{\topebases}{1}{%
\ginitdims{4}{0.4}%
\abovepic{#1}%
\present{\pretopebases{#1}}}

\newcommand{\pregdots}[6]{%
\begin{picture}(5,8.4)(0,-2.7)%
\cell{2.5}{5.7}{b}{#1}%
\cell{1.5}{2.8}{b}{#2}%
\cell{1.5}{0.2}{t}{#3}%
\cell{2.5}{-2.7}{t}{#4}%
\cell{2.7}{4.25}{l}{#5}%
\cell{2.7}{-1.25}{l}{#6}%
\qbezier(0,1.5)(2.5,9.5)(5,1.5)%
\qbezier(0,1.5)(2.5,4)(5,1.5)%
\qbezier(0,1.5)(2.5,-1)(5,1.5)%
\qbezier(0,1.5)(2.5,-6.5)(5,1.5)%
\put(2.5,5.25){\vector(0,-1){2}}%
\put(2.5,-0.25){\vector(0,-1){2}}%
\cell{2.5}{1.7}{c}{\vdots}%
\put(5,1.5){\vector(1,-4){0}}%
\put(5,1.5){\vector(4,-3){0}}%
\put(5,1.5){\vector(4,3){0}}%
\put(5,1.5){\vector(1,4){0}}%
\end{picture}}

\mcm{\gdots}{6}{%
\ginitdims{5}{8.4}%
\abovepic{#1}%
\belowpic{#4}%
\present{\pregdots{#1}{#2}{#3}{#4}{#5}{#6}}}

\newlength{\volt}
\makeatletter

\def\diagram{\m@th\leftwidth=\z@ \rightwidth=\z@ \topheight=\z@
\botheight=\z@ \setbox\@picbox\hbox\bgroup}

\def\enddiagram{\egroup\wd\@picbox\rightwidth\unitlength
\ht\@picbox\topheight\unitlength \dp\@picbox\botheight\unitlength
\hskip\leftwidth\unitlength\box\@picbox}

\def\bfig{\begin{diagram}}
\def\efig{\end{diagram}}
\newcount\wideness \newcount\leftwidth \newcount\rightwidth
\newcount\highness \newcount\topheight \newcount\botheight

\def\ratchet#1#2{\ifnum#1<#2 \global #1=#2 \fi}

\def\putbox(#1,#2)#3{%
\horsize{\wideness}{#3} \divide\wideness by 2 {\advance\wideness
by #1 \ratchet{\rightwidth}{\wideness}} {\advance\wideness by -#1
\ratchet{\leftwidth}{\wideness}} \vertsize{\highness}{#3}
\divide\highness by 2 {\advance\highness by #2
\ratchet{\topheight}{\highness}} {\advance\highness by -#2
\ratchet{\botheight}{\highness}} \put(#1,#2){\makebox(0,0){$#3$}}}

\def\putlbox(#1,#2)#3{%
\horsize{\wideness}{#3} {\advance\wideness by #1
\ratchet{\rightwidth}{\wideness}} {\ratchet{\leftwidth}{-#1}}
\vertsize{\highness}{#3} \divide\highness by 2 {\advance\highness
by #2 \ratchet{\topheight}{\highness}} {\advance\highness by -#2
\ratchet{\botheight}{\highness}}
\put(#1,#2){\makebox(0,0)[l]{$#3$}}}

\def\putrbox(#1,#2)#3{%
\horsize{\wideness}{#3} {\ratchet{\rightwidth}{#1}}
{\advance\wideness by -#1 \ratchet{\leftwidth}{\wideness}}
\vertsize{\highness}{#3} \divide\highness by 2 {\advance\highness
by #2 \ratchet{\topheight}{\highness}} {\advance\highness by -#2
\ratchet{\botheight}{\highness}}
\put(#1,#2){\makebox(0,0)[r]{$#3$}}}

\def\adjust[#1]{} 

\newcount \coefa
\newcount \coefb
\newcount \coefc
\newcount\tempcounta
\newcount\tempcountb
\newcount\tempcountc
\newcount\tempcountd
\newcount\xext
\newcount\yext
\newcount\xoff
\newcount\yoff
\newcount\gap%
\newcount\arrowtypea
\newcount\arrowtypeb
\newcount\arrowtypec
\newcount\arrowtyped
\newcount\arrowtypee
\newcount\height
\newcount\width
\newcount\xpos
\newcount\ypos
\newcount\run
\newcount\rise
\newcount\arrowlength
\newcount\halflength
\newcount\arrowtype
\newdimen\tempdimen
\newdimen\xlen
\newdimen\ylen
\newsavebox{\tempboxa}%
\newsavebox{\tempboxb}%
\newsavebox{\tempboxc}%

\newdimen\w@dth

\def\setw@dth#1#2{\setbox\z@\hbox{\m@th$#1$}\w@dth=\wd\z@
\setbox\@ne\hbox{\m@th$#2$}\ifnum\w@dth<\wd\@ne \w@dth=\wd\@ne \fi
\advance\w@dth by 1.2em}

\def\t@^#1_#2{\allowbreak\def\n@one{#1}\def\n@two{#2}\mathrel
{\setw@dth{#1}{#2} \mathop{\hbox to
\w@dth{\rightarrowfill}}\limits \ifx\n@one\empty\else
^{\box\z@}\fi \ifx\n@two\empty\else _{\box\@ne}\fi}}
\def\t@@^#1{\@ifnextchar_{\t@^{#1}}{\t@^{#1}_{}}}
\def\to{\@ifnextchar^{\t@@}{\t@@^{}}}

\def\t@left^#1_#2{\def\n@one{#1}\def\n@two{#2}\mathrel{\setw@dth{#1}{#2}
\mathop{\hbox to \w@dth{\leftarrowfill}}\limits
\ifx\n@one\empty\else ^{\box\z@}\fi \ifx\n@two\empty\else
_{\box\@ne}\fi}}
\def\t@@left^#1{\@ifnextchar_{\t@left^{#1}}{\t@left^{#1}_{}}}
\def\toleft{\@ifnextchar^{\t@@left}{\t@@left^{}}}

\def\two@^#1_#2{\allowbreak
\def\n@one{#1}\def\n@two{#2}\mathrel{\setw@dth{#1}{#2}
\mathop{\vcenter{\lineskip\z@\baselineskip\z@
                 \hbox to \w@dth{\rightarrowfill}%
                 \hbox to \w@dth{\rightarrowfill}}%
       }\limits
\ifx\n@one\empty\else ^{\box\z@}\fi \ifx\n@two\empty\else
_{\box\@ne}\fi}}
\def\tw@@^#1{\@ifnextchar _{\two@^{#1}}{\two@^{#1}_{}}}
\def\two{\@ifnextchar ^{\tw@@}{\tw@@^{}}}

\def\tofr@^#1_#2{\def\n@one{#1}\def\n@two{#2}\mathrel{\setw@dth{#1}{#2}
\mathop{\vcenter{\hbox to \w@dth{\rightarrowfill}\kern-1.7ex
                 \hbox to \w@dth{\leftarrowfill}}%
       }\limits
\ifx\n@one\empty\else ^{\box\z@}\fi \ifx\n@two\empty\else
_{\box\@ne}\fi}}
\def\t@fr@^#1{\@ifnextchar_ {\tofr@^{#1}}{\tofr@^{#1}_{}}}
\def\tofro{\@ifnextchar^ {\t@fr@}{\t@fr@^{}}}

\def\mon{\mathop{\m@th\hbox to
      14.6\P@{\lasyb\char'51\hskip-2.1\P@$\arrext$\hss
$\mathord\rightarrow$}}\limits} 
\def\leftmono{\mathrel{\m@th\hbox to
14.6\P@{$\mathord\leftarrow$\hss$\arrext$\hskip-2.1\P@\lasyb\char'50%
}}\limits} 
\mathchardef\arrext="0200       

\def\settypes(#1,#2,#3){\arrowtypea#1 \arrowtypeb#2 \arrowtypec#3}
\def\settoheight#1#2{\setbox\@tempboxa\hbox{#2}#1\ht\@tempboxa\relax}%
\def\settodepth#1#2{\setbox\@tempboxa\hbox{#2}#1\dp\@tempboxa\relax}%
\def\settokens`#1`#2`#3`#4`{%
     \def\tokena{#1}\def\tokenb{#2}\def\tokenc{#3}\def\tokend{#4}}
\def\setsqparms[#1`#2`#3`#4;#5`#6]{%
\arrowtypea #1 \arrowtypeb #2 \arrowtypec #3 \arrowtyped #4
\width #5 \height #6 }
\def\setpos(#1,#2){\xpos=#1 \ypos#2}

\def\settriparms[#1`#2`#3;#4]{\settripairparms[#1`#2`#3`1`1;#4]}%

\def\settripairparms[#1`#2`#3`#4`#5;#6]{%
\arrowtypea #1 \arrowtypeb #2 \arrowtypec #3 \arrowtyped #4
\arrowtypee #5 \width #6 \height #6 }

\def\resetparms{\settripairparms[1`1`1`1`1;500]\width 500}

\resetparms

\def\mvector(#1,#2)#3{
\put(0,0){\vector(#1,#2){#3}}%
\put(0,0){\vector(#1,#2){26}}%
}
\def\evector(#1,#2)#3{{
\arrowlength #3
\put(0,0){\vector(#1,#2){\arrowlength}}%
\advance \arrowlength by-30
\put(0,0){\vector(#1,#2){\arrowlength}}%
}}

\def\horsize#1#2{%
\settowidth{\tempdimen}{$#2$}%
#1=\tempdimen \divide #1 by\unitlength }

\def\vertsize#1#2{%
\settoheight{\tempdimen}{$#2$}%
#1=\tempdimen
\settodepth{\tempdimen}{$#2$}%
\advance #1 by\tempdimen \divide #1 by\unitlength }

\def\putvector(#1,#2)(#3,#4)#5#6{{%
\ifnum3<\arrowtype \putdashvector(#1,#2)(#3,#4)#5\arrowtype \else
\ifnum\arrowtype<-3 \putdashvector(#1,#2)(#3,#4)#5\arrowtype \else
\xpos=#1 \ypos=#2 \run=#3 \rise=#4 \arrowlength=#5 \ifnum
\arrowtype<0
    \ifnum \run=0
        \advance \ypos by-\arrowlength
    \else
        \tempcounta \arrowlength
        \multiply \tempcounta by\rise
        \divide \tempcounta by\run
        \ifnum\run>0
            \advance \xpos by\arrowlength
            \advance \ypos by\tempcounta
        \else
            \advance \xpos by-\arrowlength
            \advance \ypos by-\tempcounta
        \fi
    \fi
    \multiply \arrowtype by-1
    \multiply \rise by-1
    \multiply \run by-1
\fi \ifcase \arrowtype
\or \put(\xpos,\ypos){\vector(\run,\rise){\arrowlength}}%
\or \put(\xpos,\ypos){\mvector(\run,\rise)\arrowlength}%
\or \put(\xpos,\ypos){\evector(\run,\rise){\arrowlength}}%
\fi\fi\fi }}

\def\putsplitvector(#1,#2)#3#4{
\xpos #1 \ypos #2 \arrowtype #4 \halflength #3 \arrowlength #3
\gap 140 \advance \halflength by-\gap \divide \halflength by2
\ifnum\arrowtype>0
   \ifcase \arrowtype
   \or \put(\xpos,\ypos){\line(0,-1){\halflength}}%
       \advance\ypos by-\halflength
       \advance\ypos by-\gap
       \put(\xpos,\ypos){\vector(0,-1){\halflength}}%
   \or \put(\xpos,\ypos){\line(0,-1)\halflength}%
       \put(\xpos,\ypos){\vector(0,-1)3}%
       \advance\ypos by-\halflength
       \advance\ypos by-\gap
       \put(\xpos,\ypos){\vector(0,-1){\halflength}}%
   \or \put(\xpos,\ypos){\line(0,-1)\halflength}%
       \advance\ypos by-\halflength
       \advance\ypos by-\gap
       \put(\xpos,\ypos){\evector(0,-1){\halflength}}%
   \fi
\else \arrowtype=-\arrowtype
   \ifcase\arrowtype
   \or \advance \ypos by-\arrowlength
       \put(\xpos,\ypos){\line(0,1){\halflength}}%
       \advance\ypos by\halflength
       \advance\ypos by\gap
       \put(\xpos,\ypos){\vector(0,1){\halflength}}%
   \or \advance \ypos by-\arrowlength
       \put(\xpos,\ypos){\line(0,1)\halflength}%
       \put(\xpos,\ypos){\vector(0,1)3}%
       \advance\ypos by\halflength
       \advance\ypos by\gap
       \put(\xpos,\ypos){\vector(0,1){\halflength}}%
   \or \advance \ypos by-\arrowlength
       \put(\xpos,\ypos){\line(0,1)\halflength}%
       \advance\ypos by\halflength
       \advance\ypos by\gap
       \put(\xpos,\ypos){\evector(0,1){\halflength}}%
   \fi
\fi }

\def\putmorphism(#1)(#2,#3)[#4`#5`#6]#7#8#9{{%
\run #2 \rise #3 \ifnum\rise=0
  \puthmorphism(#1)[#4`#5`#6]{#7}{#8}#9%
\else\ifnum\run=0
  \putvmorphism(#1)[#4`#5`#6]{#7}{#8}#9%
\else
\setpos(#1)%
\arrowlength #7 \arrowtype #8 \ifnum\run=0 \else\ifnum\rise=0
\else \ifnum\run>0
    \coefa=1
\else
   \coefa=-1
\fi \ifnum\arrowtype>0
   \coefb=0
   \coefc=-1
\else
   \coefb=\coefa
   \coefc=1
   \arrowtype=-\arrowtype
\fi \width=2 \multiply \width by\run \divide \width by\rise
\ifnum \width<0  \width=-\width\fi \advance\width by60 \if l#9
\width=-\width\fi
\putbox(\xpos,\ypos){#4}
{\multiply \coefa by\arrowlength
\advance\xpos by\coefa \multiply \coefa by\rise \divide \coefa
by\run \advance \ypos by\coefa
\putbox(\xpos,\ypos){#5} }%
{\multiply \coefa by\arrowlength
\divide \coefa by2 \advance \xpos by\coefa \advance \xpos by\width
\multiply \coefa by\rise \divide \coefa by\run \advance \ypos
by\coefa
\if l#9%
   \putrbox(\xpos,\ypos){#6}%
\else\if r#9%
   \putlbox(\xpos,\ypos){#6}%
\fi\fi }%
{\multiply \rise by-\coefc
\multiply \run by-\coefc \multiply \coefb by\arrowlength \advance
\xpos by\coefb \multiply \coefb by\rise \divide \coefb by\run
\advance \ypos by\coefb \multiply \coefc by70 \advance \ypos
by\coefc \multiply \coefc by\run \divide \coefc by\rise \advance
\xpos by\coefc \multiply \coefa by140 \multiply \coefa by\run
\divide \coefa by\rise \advance \arrowlength by\coefa
\ifcase\arrowtype
\or \put(\xpos,\ypos){\vector(\run,\rise){\arrowlength}}%
\or \put(\xpos,\ypos){\mvector(\run,\rise){\arrowlength}}%
\or \put(\xpos,\ypos){\evector(\run,\rise){\arrowlength}}%
\fi}\fi\fi\fi\fi}}

\newcount\numbdashes \newcount\lengthdash \newcount\increment

\def\howmanydashes{
\numbdashes=\arrowlength \lengthdash=40 \divide\numbdashes by
\lengthdash \lengthdash=\arrowlength \divide\lengthdash by
\numbdashes
\increment=\lengthdash \multiply\lengthdash by 3
\divide\lengthdash by 5 }

\def\putdashvector(#1)(#2,#3)#4#5{%
\ifnum#3=0 \putdashhvector(#1){#4}#5 \else \ifnum#2=0
\putdashvvector(#1){#4}#5\fi\fi}

\def\putdashhvector(#1,#2)#3#4{{%
\arrowlength=#3 \howmanydashes
\multiput(#1,#2)(\increment,0){\numbdashes}%
{\vrule height .4pt width \lengthdash\unitlength} \arrowtype=#4
\xpos=#1 \ifnum\arrowtype<0 \advance\arrowtype by 7 \fi
\ifcase\arrowtype \or \advance\xpos by 10
    \put(\xpos,#2){\vector(-1,0){\lengthdash}}
    \advance\xpos by 40
    \put(\xpos,#2){\vector(-1,0){\lengthdash}}
\or \advance \xpos by 10
    \put(\xpos,#2){\vector(-1,0){\lengthdash}}
    \advance\xpos by  \arrowlength
    \advance\xpos by  -50
    \put(\xpos,#2){\vector(-1,0){\lengthdash}}
\or \advance\xpos by 10
    \put(\xpos,#2){\vector(-1,0){\lengthdash}}
\or \advance\xpos by \arrowlength
    \advance\xpos by -\lengthdash
    \put(\xpos,#2){\vector(1,0){\lengthdash}}
\or {\advance\xpos by 10
    \put(\xpos,#2){\vector(1,0){\lengthdash}}}
    \advance\xpos by \arrowlength
    \advance\xpos by -\lengthdash
    \put(\xpos,#2){\vector(1,0){\lengthdash}}
\or \advance\xpos by \arrowlength
    \advance\xpos by -\lengthdash
    \put(\xpos,#2){\vector(1,0){\lengthdash}}
    \advance\xpos by -40
    \put(\xpos,#2){\vector(1,0){\lengthdash}}
   \fi
}}

\def\putdashvvector(#1,#2)#3#4{{%
\arrowlength=#3 \howmanydashes \ypos=#2 \advance\ypos by
-\arrowlength
\multiput(#1,#2)(0,\increment){\numbdashes}%
    {\vrule width .4pt height \lengthdash\unitlength}
\arrowtype=#4 \ypos=#2 \ifnum\arrowtype<0 \advance\arrowtype by 7
\fi \ifcase\arrowtype \or \advance\ypos by \arrowlength
\advance\ypos by -40
    \put(#1,\ypos){\vector(0,1){\lengthdash}}
    \advance\ypos by -40
    \put(#1,\ypos){\vector(0,1){\lengthdash}}
\or \advance\ypos by 10
    \put(#1,\ypos){\vector(0,1){\lengthdash}}
    \advance\ypos by \arrowlength \advance\ypos by -40
    \put(#1,\ypos){\vector(0,1){\lengthdash}}
\or \advance\ypos by \arrowlength \advance\ypos by -40
    \put(#1,\ypos){\vector(0,1){\lengthdash}}
\or \advance\ypos by 10
    \put(#1,\ypos){\vector(0,-1){\lengthdash}}
\or \advance\ypos by 10
    \put(#1,\ypos){\vector(0,-1){\lengthdash}}
    \advance\ypos by \arrowlength \advance\ypos by -40
    \put(#1,\ypos){\vector(0,-1){\lengthdash}}
\or \advance\ypos by 10
    \put(#1,\ypos){\vector(0,-1){\lengthdash}}
    \advance\ypos by 40
    \put(#1,\ypos){\vector(0,-1){\lengthdash}}
\fi }}

\def\puthmorphism(#1,#2)[#3`#4`#5]#6#7#8{{%
\xpos #1 \ypos #2 \width #6 \arrowlength #6 \arrowtype=#7
\putbox(\xpos,\ypos){#3\vphantom{#4}}%
{\advance \xpos by\arrowlength
\putbox(\xpos,\ypos){\vphantom{#3}#4}}%
\horsize{\tempcounta}{#3}%
\horsize{\tempcountb}{#4}%
\divide \tempcounta by2 \divide \tempcountb by2 \advance
\tempcounta by30 \advance \tempcountb by30 \advance \xpos
by\tempcounta \advance \arrowlength by-\tempcounta \advance
\arrowlength by-\tempcountb
\putvector(\xpos,\ypos)(1,0)\arrowlength\arrowtype \divide
\arrowlength by2 \advance \xpos by\arrowlength
\vertsize{\tempcounta}{#5}%
\divide\tempcounta by2 \advance \tempcounta by20
\if a#8 %
   \advance \ypos by\tempcounta
   \putbox(\xpos,\ypos){#5}%
\else
   \advance \ypos by-\tempcounta
   \putbox(\xpos,\ypos){#5}%
\fi}}

\def\putvmorphism(#1,#2)[#3`#4`#5]#6#7#8{{%
\xpos #1 \ypos #2 \arrowlength #6 \arrowtype #7
\settowidth{\xlen}{$#5$}%
\putbox(\xpos,\ypos){#3}%
{\advance \ypos by-\arrowlength
\putbox(\xpos,\ypos){#4}}%
{\advance\arrowlength by-140 \advance \ypos by-70 \ifdim\xlen>0pt
   \if m#8%
      \putsplitvector(\xpos,\ypos)\arrowlength\arrowtype
   \else
   \putvector(\xpos,\ypos)(0,-1)\arrowlength\arrowtype
   \fi
\else
   \putvector(\xpos,\ypos)(0,-1)\arrowlength\arrowtype
\fi}%
\ifdim\xlen>0pt
   \divide \arrowlength by2
   \advance\ypos by-\arrowlength
   \if l#8%
      \advance \xpos by-40
      \putrbox(\xpos,\ypos){#5}%
   \else\if r#8%
      \advance \xpos by40
      \putlbox(\xpos,\ypos){#5}%
   \else
      \putbox(\xpos,\ypos){#5}%
   \fi\fi
\fi }}

\def\putsquarep<#1>(#2)[#3;#4`#5`#6`#7]{{%
\setsqparms[#1]%
\setpos(#2)%
\settokens`#3`%
\puthmorphism(\xpos,\ypos)[\tokenc`\tokend`{#7}]{\width}{\arrowtyped}b%
\advance\ypos by \height
\puthmorphism(\xpos,\ypos)[\tokena`\tokenb`{#4}]{\width}{\arrowtypea}a%
\putvmorphism(\xpos,\ypos)[``{#5}]{\height}{\arrowtypeb}l%
\advance\xpos by \width
\putvmorphism(\xpos,\ypos)[``{#6}]{\height}{\arrowtypec}r%
}}

\def\putsquare{\@ifnextchar <{\putsquarep}{\putsquarep%
   <\arrowtypea`\arrowtypeb`\arrowtypec`\arrowtyped;\width`\height>}}
\def\square{\@ifnextchar< {\squarep}{\squarep
   <\arrowtypea`\arrowtypeb`\arrowtypec`\arrowtyped;\width`\height>}}
\def\squarep<#1>[#2`#3`#4`#5;#6`#7`#8`#9]{{
\setsqparms[#1]
\diagram
\putsquarep<\arrowtypea`\arrowtypeb`\arrowtypec`
\arrowtyped;\width`\height>
(0,0)[#2`#3`#4`{#5};#6`#7`#8`{#9}]
\enddiagram
}}                                                 
\def\putptrianglep<#1>(#2,#3)[#4`#5`#6;#7`#8`#9]{{%
\settriparms[#1]%
\xpos=#2 \ypos=#3 \advance\ypos by \height
\puthmorphism(\xpos,\ypos)[#4`#5`{#7}]{\height}{\arrowtypea}a%
\putvmorphism(\xpos,\ypos)[`#6`{#8}]{\height}{\arrowtypeb}l%
\advance\xpos by\height
\putmorphism(\xpos,\ypos)(-1,-1)[``{#9}]{\height}{\arrowtypec}r%
}}

\def\putptriangle{\@ifnextchar <{\putptrianglep}{\putptrianglep
   <\arrowtypea`\arrowtypeb`\arrowtypec;\height>}}
\def\ptriangle{\@ifnextchar <{\ptrianglep}{\ptrianglep
   <\arrowtypea`\arrowtypeb`\arrowtypec;\height>}}
\def\ptrianglep<#1>[#2`#3`#4;#5`#6`#7]{{
\settriparms[#1]
\diagram
\putptrianglep<\arrowtypea`\arrowtypeb`
\arrowtypec;\height>
(0,0)[#2`#3`#4;#5`#6`{#7}]
\enddiagram
}}                                            

\def\putqtrianglep<#1>(#2,#3)[#4`#5`#6;#7`#8`#9]{{%
\settriparms[#1]%
\xpos=#2 \ypos=#3 \advance\ypos by\height
\puthmorphism(\xpos,\ypos)[#4`#5`{#7}]{\height}{\arrowtypea}a%
\putmorphism(\xpos,\ypos)(1,-1)[``{#8}]{\height}{\arrowtypeb}l%
\advance\xpos by\height
\putvmorphism(\xpos,\ypos)[`#6`{#9}]{\height}{\arrowtypec}r%
}}

\def\putqtriangle{\@ifnextchar <{\putqtrianglep}{\putqtrianglep
   <\arrowtypea`\arrowtypeb`\arrowtypec;\height>}}
\def\qtriangle{\@ifnextchar <{\qtrianglep}{\qtrianglep
   <\arrowtypea`\arrowtypeb`\arrowtypec;\height>}}
\def\qtrianglep<#1>[#2`#3`#4;#5`#6`#7]{{
\settriparms[#1]
\width=\height                                
\diagram
\putqtrianglep<\arrowtypea`\arrowtypeb`
\arrowtypec;\height>
(0,0)[#2`#3`#4;#5`#6`{#7}]
\enddiagram
}}

\def\putdtrianglep<#1>(#2,#3)[#4`#5`#6;#7`#8`#9]{{%
\settriparms[#1]%
\xpos=#2 \ypos=#3
\puthmorphism(\xpos,\ypos)[#5`#6`{#9}]{\height}{\arrowtypec}b%
\advance\xpos by \height \advance\ypos by\height
\putmorphism(\xpos,\ypos)(-1,-1)[``{#7}]{\height}{\arrowtypea}l%
\putvmorphism(\xpos,\ypos)[#4``{#8}]{\height}{\arrowtypeb}r%
}}

\def\putdtriangle{\@ifnextchar <{\putdtrianglep}{\putdtrianglep
   <\arrowtypea`\arrowtypeb`\arrowtypec;\height>}}
\def\dtriangle{\@ifnextchar <{\dtrianglep}{\dtrianglep
   <\arrowtypea`\arrowtypeb`\arrowtypec;\height>}}
\def\dtrianglep<#1>[#2`#3`#4;#5`#6`#7]{{
\settriparms[#1]
\width=\height                                
\diagram
\putdtrianglep<\arrowtypea`\arrowtypeb`
\arrowtypec;\height>
(0,0)[#2`#3`#4;#5`#6`{#7}]
\enddiagram
}}

\def\putbtrianglep<#1>(#2,#3)[#4`#5`#6;#7`#8`#9]{{%
\settriparms[#1]%
\xpos=#2 \ypos=#3
\puthmorphism(\xpos,\ypos)[#5`#6`{#9}]{\height}{\arrowtypec}b%
\advance\ypos by\height
\putmorphism(\xpos,\ypos)(1,-1)[``{#8}]{\height}{\arrowtypeb}r%
\putvmorphism(\xpos,\ypos)[#4``{#7}]{\height}{\arrowtypea}l%
}}

\def\putbtriangle{\@ifnextchar <{\putbtrianglep}{\putbtrianglep
   <\arrowtypea`\arrowtypeb`\arrowtypec;\height>}}
\def\btriangle{\@ifnextchar <{\btrianglep}{\btrianglep
   <\arrowtypea`\arrowtypeb`\arrowtypec;\height>}}
\def\btrianglep<#1>[#2`#3`#4;#5`#6`#7]{{
\settriparms[#1]
\width=\height                               
\diagram
\putbtrianglep<\arrowtypea`\arrowtypeb`
\arrowtypec;\height>
(0,0)[#2`#3`#4;#5`#6`{#7}]
\enddiagram
}}

\def\putAtrianglep<#1>(#2,#3)[#4`#5`#6;#7`#8`#9]{{%
\settriparms[#1]%
\xpos=#2 \ypos=#3 {\multiply \height by2
\puthmorphism(\xpos,\ypos)[#5`#6`{#9}]{\height}{\arrowtypec}b}%
\advance\xpos by\height \advance\ypos by\height
\putmorphism(\xpos,\ypos)(-1,-1)[#4``{#7}]{\height}{\arrowtypea}l%
\putmorphism(\xpos,\ypos)(1,-1)[``{#8}]{\height}{\arrowtypeb}r%
}}

\def\putAtriangle{\@ifnextchar <{\putAtrianglep}{\putAtrianglep
   <\arrowtypea`\arrowtypeb`\arrowtypec;\height>}}
\def\Atriangle{\@ifnextchar <{\Atrianglep}{\Atrianglep
   <\arrowtypea`\arrowtypeb`\arrowtypec;\height>}}
\def\Atrianglep<#1>[#2`#3`#4;#5`#6`#7]{{
\settriparms[#1]
\width=\height                                     
\diagram
\putAtrianglep<\arrowtypea`\arrowtypeb`
\arrowtypec;\height>
(0,0)[#2`#3`#4;#5`#6`{#7}]
\enddiagram
}}

\def\putAtrianglepairp<#1>(#2)[#3;#4`#5`#6`#7`#8]{{%
\settripairparms[#1]%
\setpos(#2)%
\settokens`#3`%
\puthmorphism(\xpos,\ypos)[\tokenb`\tokenc`{#7}]{\height}{\arrowtyped}b%
\advance\xpos by\height
\puthmorphism(\xpos,\ypos)[\phantom{\tokenc}`\tokend`{#8}]%
{\height}{\arrowtypee}b%
\advance\ypos by\height
\putmorphism(\xpos,\ypos)(-1,-1)[\tokena``{#4}]{\height}{\arrowtypea}l%
\putvmorphism(\xpos,\ypos)[``{#5}]{\height}{\arrowtypeb}m%
\putmorphism(\xpos,\ypos)(1,-1)[``{#6}]{\height}{\arrowtypec}r%
}}

\def\putAtrianglepair{\@ifnextchar <{\putAtrianglepairp}{\putAtrianglepairp%
   <\arrowtypea`\arrowtypeb`\arrowtypec`\arrowtyped`\arrowtypee;\height>}}
\def\Atrianglepair{\@ifnextchar <{\Atrianglepairp}{\Atrianglepairp%
   <\arrowtypea`\arrowtypeb`\arrowtypec`\arrowtyped`\arrowtypee;\height>}}

\def\Atrianglepairp<#1>[#2;#3`#4`#5`#6`#7]{{
\settripairparms[#1]
\settokens`#2`
\width=\height                                
\diagram
\putAtrianglepairp                            
<\arrowtypea`\arrowtypeb`\arrowtypec`
\arrowtyped`\arrowtypee;\height>
(0,0)[{#2};#3`#4`#5`#6`{#7}]
\enddiagram
}}

\def\putVtrianglep<#1>(#2,#3)[#4`#5`#6;#7`#8`#9]{{%
\settriparms[#1]%
\xpos=#2 \ypos=#3 \advance\ypos by\height {\multiply\height by2
\puthmorphism(\xpos,\ypos)[#4`#5`{#7}]{\height}{\arrowtypea}a}%
\putmorphism(\xpos,\ypos)(1,-1)[`#6`{#8}]{\height}{\arrowtypeb}l%
\advance\xpos by\height \advance\xpos by\height
\putmorphism(\xpos,\ypos)(-1,-1)[``{#9}]{\height}{\arrowtypec}r%
}}

\def\putVtriangle{\@ifnextchar <{\putVtrianglep}{\putVtrianglep
   <\arrowtypea`\arrowtypeb`\arrowtypec;\height>}}
\def\Vtriangle{\@ifnextchar <{\Vtrianglep}{\Vtrianglep
   <\arrowtypea`\arrowtypeb`\arrowtypec;\height>}}
\def\Vtrianglep<#1>[#2`#3`#4;#5`#6`#7]{{
\settriparms[#1]
\width=\height                                 
\diagram
\putVtrianglep<\arrowtypea`\arrowtypeb`
\arrowtypec;\height>
(0,0)[#2`#3`#4;#5`#6`{#7}]
\enddiagram
}}

\def\putVtrianglepairp<#1>(#2)[#3;#4`#5`#6`#7`#8]{{
\settripairparms[#1]%
\setpos(#2)%
\settokens`#3`%
\advance\ypos by\height
\putmorphism(\xpos,\ypos)(1,-1)[`\tokend`{#6}]{\height}{\arrowtypec}l%
\puthmorphism(\xpos,\ypos)[\tokena`\tokenb`{#4}]{\height}{\arrowtypea}a%
\advance\xpos by\height
\puthmorphism(\xpos,\ypos)[\phantom{\tokenb}`\tokenc`{#5}]%
{\height}{\arrowtypeb}a%
\putvmorphism(\xpos,\ypos)[``{#7}]{\height}{\arrowtyped}m%
\advance\xpos by\height
\putmorphism(\xpos,\ypos)(-1,-1)[``{#8}]{\height}{\arrowtypee}r%
}}

\def\putVtrianglepair{\@ifnextchar <{\putVtrianglepairp}{\putVtrianglepairp%
    <\arrowtypea`\arrowtypeb`\arrowtypec`\arrowtyped`\arrowtypee;\height>}}
\def\Vtrianglepair{\@ifnextchar <{\Vtrianglepairp}{\Vtrianglepairp%
    <\arrowtypea`\arrowtypeb`\arrowtypec`\arrowtyped`\arrowtypee;\height>}}
\def\Vtrianglepairp<#1>[#2;#3`#4`#5`#6`#7]{{
\settripairparms[#1]
\settokens`#2`
\diagram
\putVtrianglepairp                             
<\arrowtypea`\arrowtypeb`\arrowtypec`
\arrowtyped`\arrowtypee;\height>
(0,0)[{#2};#3`#4`#5`#6`{#7}]
\enddiagram
}}

\def\putCtrianglep<#1>(#2,#3)[#4`#5`#6;#7`#8`#9]{{%
\settriparms[#1]%
\xpos=#2 \ypos=#3 \advance\ypos by\height
\putmorphism(\xpos,\ypos)(1,-1)[``{#9}]{\height}{\arrowtypec}l%
\advance\xpos by\height \advance\ypos by\height
\putmorphism(\xpos,\ypos)(-1,-1)[#4`#5`{#7}]{\height}{\arrowtypea}l%
{\multiply\height by 2
\putvmorphism(\xpos,\ypos)[`#6`{#8}]{\height}{\arrowtypeb}r}%
}}

\def\putCtriangle{\@ifnextchar <{\putCtrianglep}{\putCtrianglep
    <\arrowtypea`\arrowtypeb`\arrowtypec;\height>}}
\def\Ctriangle{\@ifnextchar <{\Ctrianglep}{\Ctrianglep
    <\arrowtypea`\arrowtypeb`\arrowtypec;\height>}}
\def\Ctrianglep<#1>[#2`#3`#4;#5`#6`#7]{{
\settriparms[#1]
\width=\height                               
\diagram
\putCtrianglep<\arrowtypea`\arrowtypeb`
\arrowtypec;\height>
(0,0)[#2`#3`#4;#5`#6`{#7}]
\enddiagram
}}                                           
\def\putDtrianglep<#1>(#2,#3)[#4`#5`#6;#7`#8`#9]{{%
\settriparms[#1]%
\xpos=#2 \ypos=#3 \advance\xpos by\height \advance\ypos by\height
\putmorphism(\xpos,\ypos)(-1,-1)[``{#9}]{\height}{\arrowtypec}r%
\advance\xpos by-\height \advance\ypos by\height
\putmorphism(\xpos,\ypos)(1,-1)[`#5`{#8}]{\height}{\arrowtypeb}r%
{\multiply\height by 2
\putvmorphism(\xpos,\ypos)[#4`#6`{#7}]{\height}{\arrowtypea}l}%
}}

\def\putDtriangle{\@ifnextchar <{\putDtrianglep}{\putDtrianglep
    <\arrowtypea`\arrowtypeb`\arrowtypec;\height>}}
\def\Dtriangle{\@ifnextchar <{\Dtrianglep}{\Dtrianglep
   <\arrowtypea`\arrowtypeb`\arrowtypec;\height>}}
\def\Dtrianglep<#1>[#2`#3`#4;#5`#6`#7]{{
\settriparms[#1]
\width=\height                              
\diagram
\putDtrianglep<\arrowtypea`\arrowtypeb`
\arrowtypec;\height>
(0,0)[#2`#3`#4;#5`#6`{#7}]
\enddiagram
}}                                          
\def\setrecparms[#1`#2]{\width=#1 \height=#2}%

\def\recursep<#1`#2>[#3;#4`#5`#6`#7`#8]{{\m@th
\width=#1 \height=#2 \settokens`#3`
\settowidth{\tempdimen}{$\tokena$} \ifdim\tempdimen=0pt
  \savebox{\tempboxa}{\hbox{$\tokenb$}}%
  \savebox{\tempboxb}{\hbox{$\tokend$}}%
  \savebox{\tempboxc}{\hbox{$#6$}}%
\else
  \savebox{\tempboxa}{\hbox{$\hbox{$\tokena$}\times\hbox{$\tokenb$}$}}%
  \savebox{\tempboxb}{\hbox{$\hbox{$\tokena$}\times\hbox{$\tokend$}$}}%
  \savebox{\tempboxc}{\hbox{$\hbox{$\tokena$}\times\hbox{$#6$}$}}%
\fi \ypos=\height \divide\ypos by 2 \xpos=\ypos \advance\xpos by
\width \bfig
\putCtrianglep<-1`1`1;\ypos>(0,0)[`\tokenc`;#5`#6`{#7}]%
\puthmorphism(\ypos,0)[\tokend`\usebox{\tempboxb}`{#8}]{\width}{-1}b%
\puthmorphism(\ypos,\height)[\tokenb`\usebox{\tempboxa}`{#4}]{\width}{-1}a%
\advance\ypos by \width
\putvmorphism(\ypos,\height)[``\usebox{\tempboxc}]{\height}1r%
\efig }}

\def\recurse{\@ifnextchar <{\recursep}{\recursep<\width`\height>}}

\def\puttwohmorphisms(#1,#2)[#3`#4;#5`#6]#7#8#9{{%
\puthmorphism(#1,#2)[#3`#4`]{#7}0a \ypos=#2 \advance\ypos by 20
\puthmorphism(#1,\ypos)[\phantom{#3}`\phantom{#4}`#5]{#7}{#8}a
\advance\ypos by -40
\puthmorphism(#1,\ypos)[\phantom{#3}`\phantom{#4}`#6]{#7}{#9}b }}

\def\puttwovmorphisms(#1,#2)[#3`#4;#5`#6]#7#8#9{{%
\putvmorphism(#1,#2)[#3`#4`]{#7}0a \xpos=#1 \advance\xpos by -20
\putvmorphism(\xpos,#2)[\phantom{#3}`\phantom{#4}`#5]{#7}{#8}l
\advance\xpos by 40
\putvmorphism(\xpos,#2)[\phantom{#3}`\phantom{#4}`#6]{#7}{#9}r }}

\def\puthcoequalizer(#1)[#2`#3`#4;#5`#6`#7]#8#9{{%
\setpos(#1)%
\puttwohmorphisms(\xpos,\ypos)[#2`#3;#5`#6]{#8}11%
\advance\xpos by #8
\puthmorphism(\xpos,\ypos)[\phantom{#3}`#4`#7]{#8}1{#9} }}

\def\putvcoequalizer(#1)[#2`#3`#4;#5`#6`#7]#8#9{{%
\setpos(#1)%
\puttwovmorphisms(\xpos,\ypos)[#2`#3;#5`#6]{#8}11%
\advance\ypos by -#8
\putvmorphism(\xpos,\ypos)[\phantom{#3}`#4`#7]{#8}1{#9} }}

\def\putthreehmorphisms(#1)[#2`#3;#4`#5`#6]#7(#8)#9{{%
\setpos(#1) \settypes(#8)
\if a#9 %
     \vertsize{\tempcounta}{#5}%
     \vertsize{\tempcountb}{#6}%
     \ifnum \tempcounta<\tempcountb \tempcounta=\tempcountb \fi
\else
     \vertsize{\tempcounta}{#4}%
     \vertsize{\tempcountb}{#5}%
     \ifnum \tempcounta<\tempcountb \tempcounta=\tempcountb \fi
\fi \advance \tempcounta by 60
\puthmorphism(\xpos,\ypos)[#2`#3`#5]{#7}{\arrowtypeb}{#9}
\advance\ypos by \tempcounta
\puthmorphism(\xpos,\ypos)[\phantom{#2}`\phantom{#3}`#4]{#7}{\arrowtypea}{#9}
\advance\ypos by -\tempcounta \advance\ypos by -\tempcounta
\puthmorphism(\xpos,\ypos)[\phantom{#2}`\phantom{#3}`#6]{#7}{\arrowtypec}{#9}
}}

\def\setarrowtoks[#1`#2`#3`#4`#5`#6]{%
\def\toka{#1}
\def\tokb{#2}
\def\tokc{#3}
\def\tokd{#4}
\def\toke{#5}
\def\tokf{#6}
}
\def\hex{\@ifnextchar <{\hexp}{\hexp<1000`400>}}
\def\hexp<#1`#2>[#3`#4`#5`#6`#7`#8;#9]{%
\setarrowtoks[#9] \yext=#2 \advance \yext by #2 \xext=#1
\advance\xext by \yext \bfig
\putCtriangle<-1`0`1;#2>(0,0)[`#5`;\tokb``\tokd] \xext=#1
\yext=#2 \advance \yext by #2
\putsquare<1`0`0`1;\xext`\yext>(#2,0)[#3`#4`#7`#8;\toka```\tokf]
\advance \xext by #2
\putDtriangle<0`1`-1;#2>(\xext,0)[`#6`;`\tokc`\toke] \efig }

\makeatother

\input{tcilatex}
\begin{document}

\title{Dynamics and Control of Humanoid Robots:\\
A Geometrical Approach}
\author{Vladimir G. Ivancevic\thanks{%
Land Operation Division, Defence Science \& Technology Organisation,
Australia, Edinburgh SA 1500. Email: Vladimir.Ivancevic@dsto.defence.gov.au}
~and Tijana T. Ivancevic\thanks{%
QLIWW IP Pty Ltd. \& Tesla Science Evolution Institute, Adelaide, Australia.
Email: tivancevic@optusnet.com.au}}
\date{}
\maketitle

\begin{abstract}
This paper reviews modern geometrical dynamics and control of humanoid
robots. This general Lagrangian and Hamiltonian formalism starts with a
proper definition of humanoid's configuration manifold, which is a set of
all robot's active joint angles. Based on the `covariant force law', the
general humanoid's dynamics and control are developed. Autonomous Lagrangian
dynamics is formulated on the associated `humanoid velocity phase space',
while autonomous Hamiltonian dynamics is formulated on the associated
`humanoid momentum phase space'. Neural-like hierarchical humanoid control
naturally follows this geometrical prescription. This purely rotational and
autonomous dynamics and control is then generalized into the framework of
modern non-autonomous biomechanics, defining the Hamiltonian fitness
function. The paper concludes with several simulation examples.  \bigbreak

\noindent\textbf{Key Words:} Humanoid robots, Lagrangian and Hamiltonian
formalisms, neural-like humanoid control, time-dependent biodynamics
\end{abstract}


\section{Introduction}

Humanoid robots, being the future of robotic science, are becoming more and
more human-like in all aspects of their functioning. Both human biodynamics
and humanoid robotics are governed by Newtonian dynamical laws and
reflex--like nonlinear controls \cite%
{SIAM,IJMMS1,IJMMS2,LieLagr,GaneshSprSml}.

Although motion of humanoid robots increasingly resembles human motion, we
still need to emphasize that human joints are (and will probably always
remain) significantly more flexible than humanoid robot joints. Each joint
of a humanoid robot consists of a pair of coupled segments with only
Eulerian rotational degrees of freedom. Each human synovial joint, on the
other hand, not only exhibits gross rotational movement (roll, pitch and
yaw) but is also capable of exhibiting some hidden and restricted
translations along (X, Y, Z) axes. For example, in the knee joint, patella
(knee cap) moves for about 7--10 cm from maximal extension to maximal
flexion. It is well-known that translational amplitudes in the shoulder
joint are even greater. In other words, within the realm of rigid body
mechanics, a segment of a human arm or leg is not properly represented as a
rigid body fixed at a certain point, but rather as a rigid body hanging on
rope--like ligaments. More generally, the whole skeleton mechanically
represents a system of flexibly coupled rigid bodies, technically an
anthropomorphic topological product of SE(3)--groups. This implies more
complex kinematics, dynamics and control than in the case of humanoid robots
\cite{VladSanjeev}.

This paper reviews modern geometrical approaches to humanoid robot's
dynamics and control. It is largely based on authors' own research in
closely related fields of human biodynamics, biomechanics and humanoid
robotics. This general approach starts with a proper definition of
humanoid's configuration manifold $M$, which is a set of all active
degrees-of-freedom (DOF). Based on the \emph{covariant force law,} the
general humanoid's dynamics with large number of DOF is developed. The
tangent bundle of the manifold $M$ (called the \emph{velocity phase space})
is the stage for autonomous Lagrangian formulation of humanoid's dynamics,
while the cotangent bundle of the manifold $M$ (called the \emph{momentum
phase space}) is the stage for autonomous Hamiltonian formulation of
humanoid's dynamics. This purely rotational and autonomous robot dynamics is
then generalized along the two main lines of modern non-autonomous
biomechanics: (i) more flexible joints, and (ii) time-dependent energy
function (with energy \emph{sources} and \emph{sinks}).

In contrast to our previously published papers, the present article provides
full technical details of both autonomous and non-autonomous
(time-dependent) biodynamics and robotics, including the new \emph{%
neuro--muscular fitness dynamics.} This thorough theoretical background
would provide an interested reader with superb capability to develop their
own non-autonomous humanoid simulator.

\section{Configuration Manifold and the Covariant Force Law}

\begin{figure}[htb]
\centering \includegraphics[width=12cm]{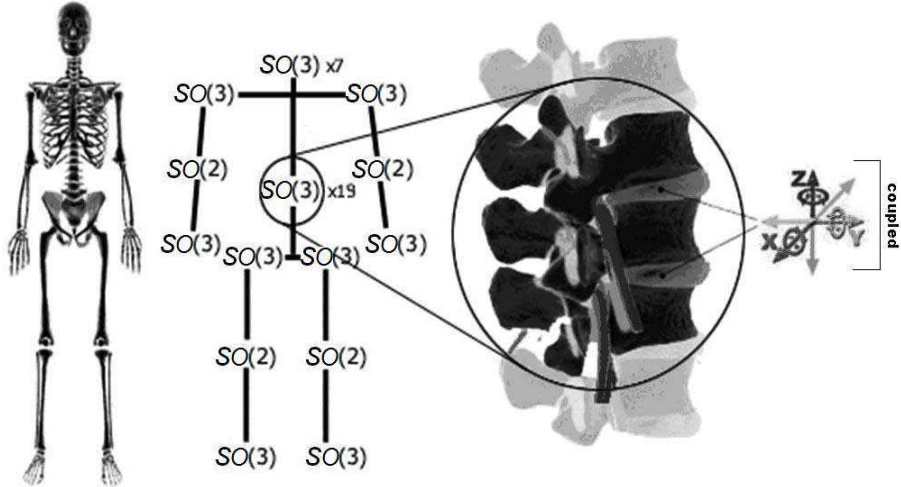}
\caption{Humanoid robot's configuration manifold $M_{rob}$, modeled upon
human skeleton. $M_{rob}$ is defined as a topological product of constrained
rotational $SO(3)$ groups, $M_{rob}=\prod_i SO(3)^i$.}
\label{SpineSO(3)}
\end{figure}

Representation of an ideal humanoid--robot motion (with human-like spine,
see Figure \ref{SpineSO(3)}) is rigorously defined in terms of \emph{%
rotational} constrained $SO(3)$--groups of motion \cite%
{Arnold,Abraham,Marsden,GaneshADG} in all main robot joints. Therefore, the
\emph{configuration manifold} $M_{rob}$ for humanoid dynamics is defined as
a topological product of all included $SO(3)$ groups, $M_{rob}=\prod_i
SO(3)^i$.

Consequently, the natural stage for autonomous Lagrangian dynamics of robot
motion is the \emph{tangent bundle} $TM_{rob}$.\footnote{%
Recall that in multibody mechanics, to each $n-$dimensional ($n$D) \textit{%
configuration manifold} $M$ there is associated its $2n$D \textit{velocity
phase--space manifold}, denoted by $TM$ and called the tangent bundle of $M$%
. The original smooth manifold $M$ is called the \emph{base} of $TM$. There
is an onto map $\pi :TM\rightarrow M$, called the \emph{projection}. Above
each point $x\in M$ there is a \textit{tangent space} $T_{x}M=\pi ^{-1}(x)$
to $M$ at $x$, which is called a \textit{fibre}. The fibre $T_{x}M\subset TM$
is the subset of $TM$, such that the total tangent bundle, $%
TM=\dbigsqcup\limits_{m\in M}T_{x}M$, is a \emph{disjoint union} of tangent
spaces $T_{x}M$ to $M$ for all points $x\in M$. From dynamical perspective,
the most important quantity in the tangent bundle concept is the smooth map $%
v:M\rightarrow TM$, which is an inverse to the projection $\pi $, i.e, $\pi
\circ v=\func{Id}_{M},\;\pi (v(x))=x$. It is called the \textit{velocity
vector--field}. Its graph $(x,v(x))$ represents the \textit{cross--section}
of the tangent bundle $TM$. This explains the dynamical term \textit{%
velocity phase--space}, given to the tangent bundle $TM$ of the manifold $M$%
. The tangent bundle is where tangent vectors live, and is itself a smooth
manifold. Vector--fields are cross-sections of the tangent bundle.} System's
\textit{Lagrangian} (energy function) is a natural energy function on the
tangent bundle \cite{LieLagr}. Similarly, the natural stage for autonomous
Hamiltonian dynamics of robot motion is the \emph{cotangent bundle} $T^{\ast
}M_{rob}$. \footnote{%
Recall that in multibody mechanics, a \emph{dual} notion to the tangent
space $T_{m}M$ to a smooth configuration manifold $M$ at a point $m$ is its
\textit{cotangent space} $T_{m}^{\ast }M$ at the same point $m$. Similarly
to the tangent bundle, for a smooth manifold $M$ of dimension $n$, its
\textit{cotangent bundle} $T^{\ast }M$ is the disjoint union of all its
cotangent spaces $T_{m}^{\ast }M$ at all points $m\in M$, i.e., $T^{\ast
}M=\dbigsqcup\limits_{m\in M}T_{m}^{\ast }M$. Therefore, the cotangent
bundle of an $n-$manifold $M$ is the vector bundle $T^{\ast }M=(TM)^{\ast }$%
, the (real) dual of the tangent bundle $TM$. The cotangent bundle is where
1--forms live, and is itself a smooth manifold. Covector--fields (1--forms)
are cross-sections of the cotangent bundle.} The \textit{Hamiltonian} is a
natural energy function on the tangent bundle \cite{IJMMS1,SIAM,IJMMS2}.

More precisely, the three--axial $SO(3)-$group of humanoid--robot joint
rotations depends on three parameters, Euler joint angles $q^{i}=(\varphi
,\psi ,\theta ),$\ defining the rotations about the Cartesian coordinate
triedar $(x,y,z)$ placed at the joint pivot point. Each of the Euler angles
are defined in the constrained range $(-\pi ,\pi )$, so the joint group
space is a constrained sphere of radius $\pi $ \cite{GaneshSprSml,GaneshADG}.

Let $G=SO(3)=\{A\in \mathcal{M}_{3\times 3}(\mathbb{R}):A^{t}A=I_{3},\det
(A)=1\}$ be the group of rotations in $\mathbb{R}^{3}$. It is a Lie group
and $\dim(G)=3$. Let us isolate its one--parameter joint subgroups, i.e.,
consider the three operators of the finite joint rotations $R_{\varphi
},R_{\psi },R_{\theta }\in SO(3),$ given by
\begin{equation*}
R_{\varphi } =\left[
\begin{array}{ccc}
1 & 0 & 0 \\
0 & \cos \varphi & -\sin \varphi \\
0 & \sin \varphi & \cos \varphi%
\end{array}
\right] , ~~ R_{\psi } =\left[
\begin{array}{ccc}
\cos \psi & 0 & \sin \psi \\
0 & 1 & 0 \\
-\sin \psi & 0 & \cos \psi%
\end{array}
\right] , ~~ R_{\theta } =\left[
\begin{array}{ccc}
\cos \theta & -\sin \theta & 0 \\
\sin \theta & \cos \theta & 0 \\
0 & 0 & 1%
\end{array}
\right]
\end{equation*}
corresponding respectively to rotations about $x-$axis by an angle $\varphi
, $ about $y-$axis by an angle $\psi ,$ and about $z-$axis by an angle $%
\theta $.

The total three--axial joint rotation $A$ is defined as the product of above
one--parameter rotations $R_{\varphi },R_{\psi },R_{\theta },$ i.e., $%
A=R_{\varphi }\cdot R_{\psi }\cdot R_{\theta }$ is equal\footnote{%
Note that this product is noncommutative, so it really depends on the order
of multiplications.}
\begin{equation*}
A=\left[
\begin{array}{ccc}
\cos \psi \cos \varphi -\cos \theta \sin \varphi \sin \psi & \cos \psi \cos
\varphi +\cos \theta \cos \varphi \sin \psi & \sin \theta \sin \psi \\
-\sin \psi \cos \varphi -\cos \theta \sin \varphi \sin \psi & -\sin \psi
\sin \varphi +\cos \theta \cos \varphi \cos \psi & \sin \theta \cos \psi \\
\sin \theta \sin \varphi & -\sin \theta \cos \varphi & \cos \theta%
\end{array}%
\right] .
\end{equation*}%
However, the order of these matrix products matters: different order
products give different results, as the matrix product is \textit{%
noncommutative product}.\footnote{%
The one--parameter rotations $R_{\varphi },R_{\psi },R_{\theta }$ define
curves in $SO(3)$ starting from $I_{3}={\small \left(
\begin{array}{ccc}
1 & 0 & 0 \\
0 & 1 & 0 \\
0 & 0 & 1%
\end{array}%
\right) }.$ Their derivatives in $\varphi =0,\psi =0$ and $\theta =0\,\ $%
belong to the associated \emph{tangent Lie algebra} $\mathfrak{so}(3)$. That
is, the corresponding infinitesimal generators of joint rotations -- joint
angular velocities $v_{\varphi },v_{\psi },v_{\theta }\in \mathfrak{so}(3)$
-- are respectively given by
\begin{eqnarray*}
v_{\varphi } &=&{\small \left[
\begin{array}{ccc}
0 & 0 & 0 \\
0 & 0 & -1 \\
0 & 1 & 0%
\end{array}%
\right] }=-y\frac{\partial }{\partial z}+z\frac{\partial }{\partial y}%
,\qquad v_{\psi }={\small \left[
\begin{array}{ccc}
0 & 0 & 1 \\
0 & 0 & 0 \\
-1 & 0 & 0%
\end{array}%
\right] }=-z\frac{\partial }{\partial x}+x\frac{\partial }{\partial z}, \\
v_{\theta } &=&{\small \left[
\begin{array}{ccc}
0 & -1 & 0 \\
1 & 1 & 0 \\
0 & 0 & 0%
\end{array}%
\right] }=-x\frac{\partial }{\partial y}+y\frac{\partial }{\partial x}.
\end{eqnarray*}%
Moreover, the elements are linearly independent and so
\begin{equation*}
\mathfrak{so}(3)=\left\{ \left[
\begin{array}{ccc}
0 & -a & b \\
a & 0 & -\gamma \\
-b & \gamma & 0%
\end{array}%
\right] |a,b,\gamma \in \mathbb{R}\right\} .
\end{equation*}%
The Lie algebra $\mathfrak{so}(3)$ is identified with $\mathbb{R}^{3}$ by
associating to each $v=(v_{\varphi },v_{\psi },v_{\theta })\in \mathbb{R}%
^{3} $ the matrix $v\in \mathfrak{so}(3)$ given by $v={\small \left[
\begin{array}{ccc}
0 & -a & b \\
a & 0 & -\gamma \\
-b & \gamma & 0%
\end{array}%
\right] }.$ Then we have the following identities:
\par
\begin{enumerate}
\item {}
\par
\item $\widehat{u\times v}=[\hat{u},v]$; and
\par
\item $u\cdot v=-\frac{1}{2}\limfunc{Tr}(\hat{u}\cdot v)$.
\end{enumerate}
\par
The exponential map $\exp :\mathfrak{so}(3)\rightarrow SO(3)$ is given by
\emph{Rodrigues relation}
\begin{equation*}
\exp (v)=I+\frac{\sin \left\Vert v\right\Vert }{\left\Vert v\right\Vert }v+%
\frac{1}{2}\left( \frac{\sin \frac{\left\Vert v\right\Vert }{2}}{\frac{%
\left\Vert v\right\Vert }{2}}\right) ^{2}v^{2}
\end{equation*}%
where the norm $\left\Vert v\right\Vert $ is given by
\begin{equation*}
\left\Vert v\right\Vert =\sqrt{(v^{1})^{2}+(v^{2})^{2}+(v^{3})^{2}}.
\end{equation*}%
\par
The the dual, \emph{cotangent Lie algebra} $\mathfrak{so}(3)^{\ast },$
includes the three joint angular momenta $p_{\varphi },p_{\psi },p_{\theta
}\in \mathfrak{so}(3)^{\ast }$, derived from the joint velocities $v$ by
multiplying them with corresponding moments of inertia.} This is the reason
why Hamilton's \textit{quaternions}\footnote{%
Recall that the set of Hamilton's \textit{quaternions} $\mathbb{H}$
represents an extension of the set of complex numbers $\mathbb{C}$. We can
compute a rotation about the unit vector, $\mathbf{u}$ by an angle $\theta $%
. The quaternion $q$ that computes this rotation is
\begin{equation*}
q=\left( \cos \frac{\theta }{2}~,~u\sin \frac{\theta }{2}\right) .
\end{equation*}%
} are today commonly used to parameterize the $SO(3)-$group, especially in
the field of 3D computer graphics.

The autonomous humanoid dynamics (both Lagrangian and Hamiltonian), is based
on the postulate of conservation of total mechanical energy. It can be
derived from the \textit{covariant force law} \cite{GaneshSprSml,GaneshADG},
which in `plain English' states:
\begin{equation*}
\text{Force 1-form}=\text{Mass distribution}\times \text{Acceleration
vector-field},
\end{equation*}%
and formally reads (using Einstein's summation convention over repeated
indices):
\begin{equation}
F_{i}=m_{ij}a^{j}.  \label{ivcov}
\end{equation}%
Here, the force 1-form $F_{i}=F_{i}(t,q,p)=F_{i}^{\prime }(t,q,\dot{q}%
),~(i=1,...,n)$ denotes any type of actuator torques; $m_{ij}$ is the
material (mass--inertia) metric tensor, which gives the total mass
distribution of the robot (including all segmental masses and their
individual inertia tensors); $a^{j}$ is the total acceleration vector-field,
including all segmental vector-fields, defined as the absolute (Bianchi)
derivative $\dot{\bar{v}}^{i}$ of all segmental angular velocities $v^{i}=%
\dot{x}^{i},~(i=1,...,n)$, where $n$ is the total number of active DOF with
local coordinates $(x^{i})$.

More formally, this \emph{central Law of robotics} represents the \textit{%
covariant force functor} $\mathcal{F}_*$ constructed over robot's
configuration manifold $M_{rob}=M$ and defined by the following commutative
diagram:

{\large
\begin{equation}
\putsquare<1`-1`-1`0;1100`500>(360,500)[TT^*M`TTM`` ;\mathcal{F}_*`F_i= \dot{%
p}_i`a^i= \dot{\bar{v}}^i`] \Vtriangle<0`-1`-1;>[T^*M=
\{x^i,p_i\}\;\;`\;\;TM= \{x^i,v^i\}`\;\;M= \{x^i\};`p_i `v^i= \dot{x}^i]
\label{covfun}
\end{equation}%
} \smallskip

The right-hand branch of the fundamental covariant force functor $\mathcal{F}%
_*:TT^*M \to TTM$ depicted in (\ref{covfun}) is Lagrangian dynamics with its
Riemannian geometry. To each $n-$dimensional ($n$D) smooth manifold $M$
there is associated its $2n$D \textit{velocity phase-space manifold},
denoted by $TM$ and called the tangent bundle of $M$. The original
configuration manifold $M$ is called the \textit{base} of $TM$. There is an
onto map $\pi :TM\rightarrow M$, called the \textit{projection}. Above each
point $x\in M$ there is a {tangent space} $T_{x}M=\pi ^{-1}(x)$ to $M$ at $x$%
, which is called a {fibre}. The fibre $T_{x}M\subset TM$ is the subset of $%
TM $, such that the total tangent bundle, $TM=\dbigsqcup\limits_{m\in
M}T_{x}M$, is a {disjoint union} of tangent spaces $T_{x}M$ to $M$ for all
points $x\in M$. From dynamical perspective, the most important quantity in
the tangent bundle concept is the smooth map $v:M\rightarrow TM$, which is
an inverse to the projection $\pi $, i.e, $\pi \circ v=\func{Id}_{M},\;\pi
(v(x))=x$. It is called the \textit{velocity vector-field} $v^i= \dot{x}^i$.%
\footnote{%
This explains the dynamical term \textit{velocity phase--space}, given to
the tangent bundle $TM$ of the manifold $M$.} Its graph $(x,v(x))$
represents the {cross--section} of the tangent bundle $TM$. Velocity
vector-fields are cross-sections of the tangent bundle. Biomechanical \emph{%
Lagrangian} (that is, kinetic minus potential energy) is a natural energy
function on the tangent bundle $TM$. The tangent bundle is itself a smooth
manifold. It has its own tangent bundle, $TTM$. Cross-sections of the second
tangent bundle $TTM$ are the acceleration vector-fields.

The left-hand branch of the fundamental covariant force functor $\mathcal{F}%
_*:TT^*M \to TTM$ depicted in (\ref{covfun}) is Hamiltonian dynamics with
its symplectic geometry. It takes place in the \textit{cotangent bundle} $%
T^{\ast }M_{rob}$, defined as follows. A \textit{dual} notion to the tangent
space $T_{x}M$ to a smooth manifold $M$ at a point $x=(x^i)$ with local is
its {cotangent space} $T_{x}^{\ast }M$ at the same point $x$. Similarly to
the tangent bundle $TM$, for any smooth $n$D manifold $M$, there is
associated its $2n$D \emph{momentum phase-space manifold}, denoted by $%
T^{\ast }M$ and called the \textit{cotangent bundle}. $T^{\ast }M$ is the
disjoint union of all its cotangent spaces $T_{x}^{\ast }M$ at all points $%
x\in M$, i.e., $T^{\ast }M=\dbigsqcup\limits_{x\in M}T_{x}^{\ast }M$.
Therefore, the cotangent bundle of an $n-$manifold $M$ is the vector bundle $%
T^{\ast }M=(TM)^{\ast }$, the (real) dual of the tangent bundle $TM$.
Momentum 1--forms (or, covector-fields) $p_i$ are cross-sections of the
cotangent bundle. Biomechanical \emph{Hamiltonian} (that is, kinetic plus
potential energy) is a natural energy function on the cotangent bundle. The
cotangent bundle $T^*M$ is itself a smooth manifold. It has its own tangent
bundle, $TT^*M$. Cross-sections of the mixed-second bundle $TT^*M$ are the
force 1--forms $F_i= \dot{p}_i$.

There is a unique smooth map from the right-hand branch to the left-hand
branch of the diagram (\ref{covfun}):
\begin{equation*}
TM\ni(x^i,v^i)\mapsto (x^i,p^i)\in T^{\ast }M.
\end{equation*}
It is called the \emph{Legendre transformation}, or \emph{fiber derivative}
(for details see, e.g. \cite{GaneshADG}).

The fundamental covariant force functor $\mathcal{F}_*:TT^*M \to TTM$ states
that the force 1--form $F_i= \dot{p}_i$, defined on the mixed
tangent--cotangent bundle $TT^*M$, causes the acceleration vector-field $%
a^i= \dot{\bar{v}}^i$, defined on the second tangent bundle $TTM$ of the
configuration manifold $M$. The corresponding \textit{contravariant
acceleration functor} is defined as its inverse map, $\mathcal{F}^*:TTM\to
TT^*M$.

\section{Lagrangian vs. Hamiltonian Approach to Humanoid Robotics}

The humanoid's configuration manifold $M_{rob}=M$ is coordinated by local
joint angular coordinates $x^i(t),~~i=1,...,n=$ total number of active DOF.
The corresponding joint angular velocities $\dot{x}^i(t)$ live in the \emph{%
velocity phase space} $TM$ (the \emph{tangent bundle} of the configuration
manifold $M$),\footnote{%
On the velocity phase--space manifold $TM$ exists:
\par
\begin{enumerate}
\item A unique $1-$form $\theta _{L}$, defined in local coordinates $%
q^{i},\,v^{i}=\dot{q}^{i}\in U_{v}$ ($U_{v}$ open in $TM$) by $\theta
_{L}\,=\,L_{v^{i}}dq^{i}$, where $L_{v^{i}}\,\equiv \,{\partial L}/{\partial
v^{i}}$.
\par
\item A unique nondegenerate Lagrangian symplectic $2-$form $\omega _{L}$,
which is closed ($d\omega _{L}\,=\,0$) and exact ($\omega _{L}\,=\,d\theta
_{L}\,=\,dL_{v^{i}}\wedge dq^{i}$).
\end{enumerate}
\par
$TM$ is an orientable manifold, admitting the standard volume given by%
\begin{equation*}
\Omega _{\omega _{L}}={\frac{{(-1)^{{\frac{{N(N+1)}}{{2}}}}}}{{N!}}}\omega
_{L}^{N},
\end{equation*}%
in local coordinates $q^{i},\,v^{i}=\dot{q}^{i}\in U_{v}$ ($U_{v}$ open in $%
TM$) it is given by
\begin{equation*}
\Omega _{L}\,=\,dq^{1}\wedge \dots \wedge dq^{N}\wedge dv^{1}\wedge \dots
\wedge dv^{N}.
\end{equation*}%
} which has the Riemannian geometry with the \textit{local metric form}:
\begin{equation*}
\langle g\rangle\equiv ds^{2}=g_{ij}dx^{i}dx^{j},\qquad\text{where}\qquad
g_{ij}(x)\,=\,m_{\mu }\delta _{rs}\frac{\partial x^{r}}{\partial q^{i}}\frac{%
\partial x^{s}}{\partial q^{j}}
\end{equation*}
is the material metric tensor defined by humanoid's \emph{mass-inertia matrix%
} (composed of individual segmental masses $m_{\mu }$) and $dx^{i}$ are
differentials of the local joint coordinates $x^i$ on $M$. Besides giving
the local distances between the points on the manifold $M,$ the Riemannian
metric form $\langle g\rangle$ defines the system's kinetic energy:
\begin{equation*}
T=\frac{1}{2}g_{ij}\dot{x}^{i}\dot{x}^{j},
\end{equation*}
giving the \emph{Lagrangian equations} of the conservative skeleton motion
with kinetic-minus-potential energy Lagrangian $L=T-V$, with the
corresponding \emph{geodesic form} \cite{GaneshADG}
\begin{equation}
\frac{d}{dt}L_{\dot{x}^{i}}-L_{x^{i}}=0\qquad\text{or,~~equivalently}\qquad
\ddot{x}^i+\Gamma _{jk}^{i}\dot{x}^{j}\dot{x}^{k}=0,  \label{geodes}
\end{equation}%
where subscripts denote partial derivatives, while $\Gamma _{jk}^{i}$ are
the Christoffel symbols of the affine Levi-Civita connection of the humanoid
manifold $M$, given by
\begin{equation*}
\Gamma _{jk}^{i}\,=\,g^{il}\Gamma _{jkl},\qquad \Gamma _{ijk}=\frac{1}{2}%
(\partial _{x^{i}}g_{jk}+\partial _{x^{j}}g_{ki}+\partial _{x^{k}}g_{ij}).
\end{equation*}

The general form of autonomous Lagrangian humanoid robotics on the
corresponding Riemannian tangent bundles $TM_{rob}$ and $TM_{hum}$ of the
configuration manifolds $M_{rob}$ and $M_{hum}$ (precisely derived in \cite%
{LieLagr}) can be formulated in a unified form as:
\begin{equation}
\frac{d}{dt}L_{\dot{x}^{i}}-L_{x^{i}}=F_{i}\left(t,x,\dot{x} \right),\qquad
(i=1,...,n),  \label{classic}
\end{equation}
where $F_{i}$ are all possible torque 1-forms, including robot's actuators,
joint dissipations and external disturbances.

On the other hand, we develop the autonomous Hamiltonian robotics on
humanoid's configuration manifold $M_{rob}=M$ in three steps, following the
standard symplectic geometry prescription (see \cite{GaneshSprSml,GaneshADG}%
):\newline

\noindent \textbf{Step A} Find a symplectic \emph{momentum phase--space} $%
(P,\omega )$.

Recall that a symplectic structure on a smooth manifold $M$ is a
nondegenerate closed\footnote{%
A $p-$form $\beta $ on a smooth manifold $M$ is called \textit{closed} if
its exterior derivative $d=\partial_i dx^i$ is equal to zero,
\begin{equation*}
d\beta=0.
\end{equation*}
From this condition one can see that the closed form (the \textit{kernel} of
the exterior derivative operator $d$) is conserved quantity. Therefore,
closed $p-$forms possess certain invariant properties, physically
corresponding to the \emph{conservation laws}.
\par
Also, a $p-$form $\beta$ that is an exterior derivative of some $(p-1)-$form
$\alpha$,
\begin{equation*}
\beta=d\alpha,
\end{equation*}
is called \textit{exact} (the \textit{image} of the exterior derivative
operator $d$). By \emph{Poincar\'e lemma,} exact forms prove to be closed
automatically,
\begin{equation*}
d\beta=d(d\alpha)=0.
\end{equation*}%
\par
Since $d^{2}=0$, \emph{every exact form is closed.} The converse is only
partially true, by Poincar\'{e} lemma: every closed form is \textit{locally
exact}.
\par
Technically, this means that given a closed $p-$form $\alpha \in \Omega
^{p}(U)$, defined on an open set $U$ of a smooth manifold $M$ any point $%
m\in U$ has a neighborhood on which there exists a $(p-1)-$form $\beta \in
\Omega ^{p-1}(U)$ such that $d\beta =\alpha |_{U}.$ In particular, there is
a Poincar\'{e} lemma for contractible manifolds: Any closed form on a
smoothly contractible manifold is exact.} $2-$form $\omega $ on $M$, i.e.,
for each $x\in M$, $\omega (x)$ is nondegenerate, and $d\omega =0$.

Let $T_{x}^*M$ be a cotangent space to $M$ at $m$. The cotangent bundle $%
T^*M $ represents a union $\cup _{m\in M}T_{x}^*M$, together with the
standard topology on $T^*M$ and a natural smooth manifold structure, the
dimension of which is twice the dimension of $M$. A $1-$form $\theta $ on $M$
represents a section $\theta :M\rightarrow T^*M$ of the cotangent bundle $%
T^*M$.

$P=T^*M$ is our momentum phase--space. On $P$ there is a nondegenerate
symplectic $2-$form $\omega $ is defined in local joint coordinates $%
x^{i},p_{i}\in U$, $U$ open in $P$, as $\omega =dx^{i}\wedge dp_{i}$. In
that case the coordinates $x^{i},p_{i}\in U$ are called canonical. In a
usual procedure the canonical $1-$form $\theta $ is first defined as $\theta
=p_{i}dx^{i}$, and then the canonical 2--form $\omega $ is defined as $%
\omega =-d\theta $.

A \emph{symplectic phase--space manifold} is a pair $(P,\omega )$.\newline

\noindent \textbf{Step B} Find a \emph{Hamiltonian vector-field} $X_H$ on $%
(P,\omega)$.

Let $(P,\omega )$ be a symplectic manifold. A vector-field $X:P\rightarrow TP
$ is called \emph{Hamiltonian} if there is a smooth function $F:P\rightarrow
\mathbb{R}$ such that $X\rfloor \omega \,=\,dF$ ($X\rfloor \omega $ ${\equiv
}$ $i_{X}\omega $ denotes the \emph{interior product} or \emph{contraction}
of the vector-field $X$ and the 2--form $\omega $). $X$ is \emph{locally
Hamiltonian} if $X\rfloor \omega $ is closed.

Let the smooth real--valued \emph{Hamiltonian function} $H:P\rightarrow
\mathbb{R}$, representing the total humanoid energy $H(x,p)\,=\,T(p)\,+\,V(x)
$ ($T$ and $V$ denote kinetic and potential energy of the system,
respectively), be given in local canonical coordinates $x^{i},p_{i}\in U$, $U
$ open in $P$. The \emph{Hamiltonian vector-field} $X_{H}$, condition by $%
X_{H}\rfloor \omega \,=\,dH$, is actually defined via symplectic matrix $J$,
in a local chart $U$, as
\begin{equation}
X_{H}=J\nabla H=\left( \partial _{p_{i}}H,-\partial _{x^{i}}H\right) ,\qquad
J={\small \left(
\begin{matrix}
0 & I \\
-I & 0%
\end{matrix}%
\right) },  \label{HamVec}
\end{equation}%
where $I$ denotes the $n\times n$ identity matrix and $\nabla $ is the
gradient operator.\newline

\noindent \textbf{Step C} Find a \emph{Hamiltonian phase--flow} $\phi _{t}$
of $X_{H}$.

Let $(P,\omega )$ be a symplectic phase--space manifold and $%
X_{H}\,=\,J\nabla H$ a Hamiltonian vector-field corresponding to a smooth
real--valued Hamiltonian function $H:P\rightarrow \mathbb{R}$, on it. If a
unique one--parameter group of diffeomorphisms $\phi _{t}:P\rightarrow P$
exists so that $\frac{d}{dt}|_{t=0}\,\phi _{t}x=J\nabla H(x)$, it is called
the \emph{Hamiltonian phase--flow}.

A smooth curve $t\mapsto \left( x^{i}(t),\,p_{i}(t)\right) $ on $(P,\omega )$
represents an \emph{integral curve} of the Hamiltonian vector-field $%
X_{H}=J\nabla H$, if in the local canonical coordinates $x^{i},p_{i}\in U$, $%
U$ open in $P$, \emph{Hamiltonian canonical equations} hold (with $\partial
_{u}\equiv \partial /\partial u$, ):
\begin{equation}
\dot{q}^{i}=\partial _{p_{i}}H,\qquad \dot{p}_{i}=-\partial _{x^{i}}H.
\label{HamEq}
\end{equation}

An integral curve is said to be \emph{maximal} if it is not a restriction of
an integral curve defined on a larger interval of $\mathbb{R}$. It follows
from the standard theorem on the \emph{existence} and \emph{uniqueness} of
the solution of a system of ODEs with smooth r.h.s, that if the manifold $%
(P,\omega )$ is Hausdorff, then for any point $x=(x^{i},p_{i})\in U$, $U$
open in $P$, there exists a maximal integral curve of $X_{H}\,=\,J\nabla H$,
passing for $t=0$, through point $x$. In case $X_{H}$ is complete, i.e., $%
X_{H}$ is $C^{p}$ and $(P,\omega )$ is compact, the maximal integral curve
of $X_{H}$ is the Hamiltonian phase--flow $\phi _{t}:U\rightarrow U$.

The phase--flow $\phi _{t}$ is \emph{symplectic} if $\omega $ is constant
along $\phi _{t}$, i.e., $\phi _{t}^*\omega =\omega$

($\phi _{t}^*\omega $ denotes the \emph{pull--back}\footnote{%
Given a map $f:X\to X^{\prime}$ between the two manifolds, the \emph{pullback%
} on $X$ of a form $\alpha$ on $X^{\prime}$ by $f$ is denoted by $f^*\alpha$%
. The pullback satisfies the relations
\begin{eqnarray*}
f^*(\alpha\wedge\beta) =f^*\alpha\wedge f^*\beta, \qquad df^*\alpha
=f^*(d\alpha),
\end{eqnarray*}
for any two forms $\alpha,\beta\in\mathbf{\Omega}^p(X)$.} of $\omega $ by $%
\phi _{t}$),

iff $\mathfrak{L}_{X_{H}}\omega \,=0$

($\mathfrak{L}_{X_{H}}\omega $ denotes the \emph{Lie derivative}\footnote{%
The \textit{Lie derivative} $\mathfrak{L}_u\alpha$ of $p-$form $\alpha$
along a vector-field $u$ is defined by Cartan's `magic' formula (see \cite%
{GaneshADG}):
\begin{equation*}
\mathfrak{L}_u\alpha =u\rfloor d\alpha +d(u\rfloor\alpha).
\end{equation*}
It satisfies the \emph{Leibnitz relation}
\begin{equation*}
\mathfrak{L}_u(\alpha\wedge\beta)= \mathfrak{L}_u\alpha\wedge\beta
+\alpha\wedge\mathfrak{L}_u\beta.
\end{equation*}
Here, the \emph{contraction} $\rfloor$ of a vector-field $u =
u^\mu\partial_\mu $ and a $p-$form $\alpha =\alpha_{\lambda_1\dots\lambda_p}
dx^{\lambda_1}\wedge\cdots\wedge dx^{\lambda_p}$ on a humanoid manifold $X$
is given in local coordinates on $X$ by
\begin{equation*}
u\rfloor\alpha = u^\mu \alpha_{\mu\lambda_1\ldots\lambda_{p-1}}
dx^{\lambda_1}\wedge\cdots \wedge dx^{\lambda_{p-1}}.
\end{equation*}
It satisfies the following relation
\begin{equation*}
u\rfloor(\alpha\wedge\beta)= u\rfloor\alpha\wedge\beta
+(-1)^{|\alpha|}\alpha\wedge u\rfloor\beta.
\end{equation*}%
} of $\omega $ upon $X_{H}$).

Symplectic phase--flow $\phi _{t}$ consists of canonical transformations on $%
(P,\omega )$, i.e., diffeomorphisms in canonical coordinates $x^{i},p_{i}\in
U$, $U$ open on all $(P,\omega )$ which leave $\omega $ invariant. In this
case the \emph{Liouville theorem} is valid: $\phi _{t}$ \emph{preserves} the
\emph{phase volume} on $(P,\omega )$. Also, the system's total energy $H$ is
conserved along $\phi _{t}$, i.e., $H\circ \phi _{t}=\phi _{t}$.

Recall that the Riemannian metrics $g\,=\,<,>$ on the configuration manifold
$M$ is a positive--definite quadratic form $g:TM\rightarrow \mathbb{R}$, in
local coordinates $x^{i}\in U$, $U$ open in $M$. Given the metrics $g_{ij}$,
the system's Hamiltonian function represents a momentum $p$--dependent
quadratic form $H:T^*M\rightarrow \mathbb{R}$ -- the system's kinetic energy
$H(p)\,=T(p)=\,\frac{1}{2}<p,p>$, in local canonical coordinates $%
x^{i},p_{i}\in U_{p}$, $U_{p}$ open in $T^*M$, given by
\begin{equation}
H(p)=\frac{1}{2}g^{ij}(x,m)\,p_{i}p_{j},  \label{dd19}
\end{equation}
where $g^{ij}(x,m)=g_{ij}^{-1}(x,m)$ denotes the \emph{inverse}
(contravariant) material \emph{metric tensor}
\begin{equation*}
g^{ij}(x,m)\,=\sum_{\chi =1}^{n}m_{\chi }\delta _{rs}\frac{\partial x^{i}}{%
\partial x^{r}}\frac{\partial x^{j}}{\partial x^{s}}.
\end{equation*}

$T^*M$ is an \emph{orientable} manifold, admitting the standard \emph{volume
form}
\begin{equation*}
\Omega _{\omega _{H}}=\,{\frac{{(-1)^{{\frac{{N(N+1)}}{{2}}}}}}{{N!}}}\omega
_{H}^{N}.
\end{equation*}

For Hamiltonian vector-field, $X_{H}$ on $M$, there is a base integral curve
$\gamma _{0}(t)\,=\,\left( x^{i}(t),\,p_{i}(t)\right) $ iff $\gamma _{0}(t)$
is a \emph{geodesic}, given by the one--form \emph{force equation}
\begin{equation}  \label{ddmom}
\dot{\bar{p}}_{i}\equiv \dot{p}_{i}+\Gamma
_{jk}^{i}\,g^{jl}g^{km}\,p_{l}p_{m}=0,\qquad \text{with \qquad }\dot{x}%
^{k}=g^{ki}p_{i}.
\end{equation}

The l.h.s $\dot{\bar{p}}_{i}$ of the covariant momentum equation (\ref{ddmom}%
)\ represents the {intrinsic} or {Bianchi covariant derivative} of the
momentum with respect to time $t$. Basic relation $\dot{\bar{p}}_{i}\,=\,0$
defines the \textit{parallel transport} on $T^{N}$, the simplest form of
humanoid's dynamics. In that case Hamiltonian vector-field $X_{H}$ is called
the \emph{geodesic spray} and its phase--flow is called the \emph{geodesic
flow}.

For Earthly dynamics in the gravitational \emph{potential} field $%
V:M\rightarrow \mathbb{R}$, the Hamiltonian $H:T^*M\rightarrow \mathbb{R}$ (%
\ref{dd19}) extends into potential form
\begin{equation*}
H(p,x)=\frac{1}{2}g^{ij}p_{i}p_{j}+V(x),
\end{equation*}
with Hamiltonian vector-field $X_{H}\,=\,J\nabla H$ still defined by
canonical equations (\ref{HamEq}).

A general form of a \emph{driven}, non--conservative Hamiltonian equations
reads:
\begin{equation}
\dot{x}^{i}=\partial _{p_{i}}H,\qquad \dot{p}_{i}=F_{i}-\partial _{x^{i}}H,
\label{FHam}
\end{equation}%
where $F_{i}=F_{i}(t,x,p)$ represent any kind of joint--driving \emph{%
covariant torques}, including active neuro--muscular--like controls, as
functions of time, angles and momenta, as well as passive dissipative and
elastic joint torques. In the covariant momentum formulation (\ref{ddmom}),
the non--conservative Hamiltonian equations (\ref{FHam}) become
\begin{equation*}
\dot{\bar{p}}_{i}\equiv \dot{p}_{i}+\Gamma
_{jk}^{i}\,g^{jl}g^{km}\,p_{l}p_{m}=F_{i}\qquad \text{with}\qquad \dot{x}%
^{k}=g^{ki}p_{i}.
\end{equation*}

The general form of autonomous Hamiltonian robotics is given by dissipative,
driven Hamiltonian equations on $T^{\ast }M$:
\begin{eqnarray}
\dot{x}^{i} &=&\partial _{p_{i}}H+\partial _{p_{i}}R,  \label{fh1} \\
\dot{p}_{i} &=&F_{i}-\partial _{x_{i}}H+\partial _{x_{i}}R,  \label{fh2} \\
x^{i}(0) &=&x_{0}^{i},\qquad p_{i}(0)=p_{i}^{0},  \label{fh3}
\end{eqnarray}%
including \textit{contravariant equation} (\ref{fh1}) -- the \textit{%
velocity vector-field}, and \textit{covariant equation} (\ref{fh2}) -- the
\textit{force 1--form} (field), together with initial joint angles and
momenta (\ref{fh3}). Here $R=R(x,p)$ denotes the Raileigh nonlinear
(biquadratic) dissipation function, and $F_{i}=F_{i}(t,x,p)$ are covariant
driving torques of robot's actuators. The velocity vector-field (\ref{fh1})
and the force $1-$form (\ref{fh2}) together define the generalized
Hamiltonian vector-field $X_{H}$; the Hamiltonian energy function $H=H(x,p)$
is its generating function.

As a Lie group, the humanoid's configuration manifold $M=\prod_{j}SO(3)^{j}$
is Hausdorff.\footnote{%
That is, for every pair of points $x_{1},x_{2}\in M$, there are disjoint
open subsets (charts) $U_{1},U_{2}\subset M$ such that $x_{1}\in U_{1}$ and $%
x_{2}\in U_{2}$.} Therefore, for $x\,=\,(x^{i},\,p_{i})\in U_{p}$, where $%
U_{p}$ is an open coordinate chart in $T^*M$, there exists a unique
one--parameter group of diffeomorphisms $\phi_t:T^*M\rightarrow T^*M$, that
is the \emph{autonomous Hamiltonian phase--flow:}
\begin{eqnarray}  \label{fh4}
\phi_t &:&T^*M\rightarrow T^*M:(p(0),x(0))\mapsto (p(t),x(t)), \\
(\phi_t &\circ & \phi _s\, =\,\phi_{t+s},\quad \phi_0\,=\,\text{identity}),
\notag
\end{eqnarray}
given by (\ref{fh1}--\ref{fh3}) such that
\begin{equation*}
{\frac{{d}}{{dt}}}\left\vert _{t=0}\right. \phi_tx\,=\,J\nabla H(x).
\end{equation*}

The general form of Hamiltonian humanoid robotics on the symplectic
cotangent bundle $T^{\ast }M_{rob}$ of the configuration manifold $M_{rob}$
(as derived in \cite{IJMMS2,VladSanjeev,VladTij}) is based on the \textit{%
affine Hamiltonian function} $H_{a}:T^{\ast }M\rightarrow \mathbb{R},$ in
local canonical coordinates on $T^{\ast }M$ given by
\begin{equation}
H_{a}(x,p,u)=H_{0}(x,p)-H^{j}(x,p)\,u_{j},  \label{aff}
\end{equation}%
where $H_{0}(x,p)$ is the physical Hamiltonian (kinetic + potential energy)
dependent on joint coordinates $x^{i}$ and canonical momenta $p^{i}$, $%
H^{j}=H^{j}(x,p)$, ($j=1,\dots ,\,m\leq n$ are the coupling Hamiltonians
corresponding to the system's active joints and $u_{i}=u_{i}(t,x,p)$ are
(reflex) feedback--controls. Using (\ref{aff}) we come to the affine
Hamiltonian control HBE--system, in deterministic form%
\begin{align}
\dot{x}^{i}& =\partial _{p_{i}}H_{0}-\partial _{p_{i}}H^{j}\,u_{j}+\partial
_{p_{i}}R,  \label{af1} \\
\dot{p}_{i}& =\mathcal{F}_{i}(t,x,p)-\partial _{x^{i}}H_{0}+\partial
_{x^{i}}H^{j}\,u_{j}+\partial _{x^{i}}R,  \notag \\
o^{i}& =-\partial _{u_{i}}H_{a}=H^{j},  \notag \\
x^{i}(0)& =x_{0}^{i},\qquad p_{i}(0)=p_{i}^{0},  \notag \\
(i& =1,\dots ,n;\qquad j=1,\dots ,\,M\leq n),  \notag
\end{align}%
($\mathcal{F}_{i}=\mathcal{F}_{i}(t,x,p),$ $H_{0}=H_{0}(x,p),$ $%
H^{j}=H^{j}(x,p),$ $H_{a}=H_{a}(x,p,u),$ $R=R(x,p)$), as well as in the
fuzzy--stochastic form%
\begin{align}
dq^{i}& =\left( \partial _{p_{i}}H_{0}(\sigma _{\mu })-\partial
_{p_{i}}H^{j}(\sigma _{\mu })\,u_{j}+\partial _{p_{i}}R\right) \,dt,  \notag
\\
dp_{i}& =B_{ij}[x^{i}(t),t]\,dW^{j}(t)\qquad +\qquad \qquad   \label{af2} \\
& \left( \bar{\mathcal{F}}_{i}(t,x,p)-\partial _{x^{i}}H_{0}(\sigma _{\mu
})+\partial _{x^{i}}H^{j}(\sigma _{\mu })\,u_{j}+\partial _{x^{i}}R\right)
\,dt,  \notag \\
d\bar{o}^{i}& =-\partial _{u_{i}}H_{a}(\sigma _{\mu })\,dt=H^{j}(\sigma
_{\mu })\,dt,\qquad \qquad   \notag \\
x^{i}(0)& =\bar{x}_{0}^{i},\qquad p_{i}(0)=\bar{p}_{i}^{0}\qquad \qquad
\notag
\end{align}%
In (\ref{af1})--(\ref{af2}), $R=R(x,p)$ denotes the joint (nonlinear)
dissipation function, $o^{i}$ are affine system outputs (which can be
different from joint coordinates); $\{\sigma \}_{\mu }$ \ (with $\mu \geq 1$%
) denote fuzzy sets of conservative parameters (segment lengths, masses and
moments of inertia), dissipative joint dampings and actuator parameters
(amplitudes and frequencies), while the bar $\bar{(.)}$ over a variable
denotes the corresponding fuzzified variable; $B_{ij}[q^{i}(t),t]$ denote
diffusion fluctuations and $W^{j}(t)$ are discontinuous jumps as the $n$%
--dimensional Wiener process.

\section{Generalization to Human Biodynamics}

If we neglect anatomy and physiology of human sensors and effectors, that
is, from purely mechanical perspective, there are two main dynamical
differences between robots and humans: (i) human joints are more flexible
than robot joints (effectively many more degrees-of-freedom), and (ii) human
dynamics is usually non-autonomous, or time-dependent. We will explain both
differences in some detail in the following subsections.

\subsection{Realistic Configuration Manifold of Human Motion}

Every rotation in all synovial human joints is followed by the corresponding
micro--translation, which occurs after the rotational amplitude is reached
\cite{VladSanjeev}. So, representation of human motion is rigorously defined
in terms of \emph{Euclidean} $SE(3)$--groups of full rigid--body motion \cite%
{Marsden,GaneshSprSml,GaneshADG} in all main human joints (see Figure \ref%
{SpineSE(3)}). Therefore, the configuration manifold $M_{hum}$ for human
dynamics is defined as a topological product of all included constrained $%
SE(3)$ groups, $M_{hum}=\prod_i SE(3)^i$. Consequently, the natural stage
for autonomous Lagrangian dynamics of human motion is the tangent bundle $%
TM_{hum}$ \cite{LieLagr}, and for the corresponding autonomous Hamiltonian
dynamics is the cotangent bundle $T^*M_{hum}$ \cite{IJMMS1,SIAM,IJMMS2}.
\begin{figure}[htb]
\centering \includegraphics[width=12cm]{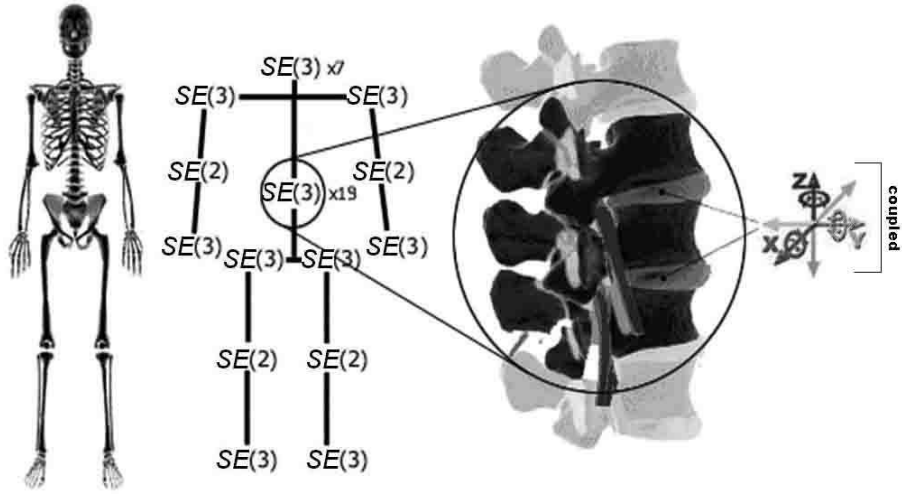}
\caption{The configuration manifold $M_{hum}$ of the human body is defined
as a topological product of constrained $SE(3)$ groups acting in all major
(synovial) human joints, $M_{hum}=\prod_i SE(3)^i$.}
\label{SpineSE(3)}
\end{figure}

Briefly, the Euclidean SE(3)--group is defined as a semidirect
(noncommutative) product of 3D rotations and 3D translations, $%
SE(3):=SO(3)\rhd \mathbb{R}^{3}$. Its most important subgroups are the
following \cite{GaneshSprBig,ParkChung,GaneshADG}):\newline

{{\frame{$%
\begin{array}{cc}
\mathbf{Subgroup} & \mathbf{Definition} \\ \hline
\begin{array}{c}
SO(3),\text{ group of rotations} \\
\text{in 3D (a spherical joint)}%
\end{array}
&
\begin{array}{c}
\text{Set of all proper orthogonal } \\
3\times 3-\text{rotational matrices}%
\end{array}
\\ \hline
\begin{array}{c}
SE(2),\text{ special Euclidean group} \\
\text{in 2D (all planar motions)}%
\end{array}
&
\begin{array}{c}
\text{Set of all }3\times 3-\text{matrices:} \\
\left[
\begin{array}{ccc}
\cos \theta & \sin \theta & r_{x} \\
-\sin \theta & \cos \theta & r_{y} \\
0 & 0 & 1%
\end{array}%
\right]%
\end{array}
\\ \hline
\begin{array}{c}
SO(2),\text{ group of rotations in 2D} \\
\text{subgroup of }SE(2)\text{--group} \\
\text{(a revolute joint)}%
\end{array}
&
\begin{array}{c}
\text{Set of all proper orthogonal } \\
2\times 2-\text{rotational matrices} \\
\text{ included in }SE(2)-\text{group}%
\end{array}
\\ \hline
\begin{array}{c}
\mathbb{R}^{3},\text{ group of translations in 3D} \\
\text{(all spatial displacements)}%
\end{array}
& \text{Euclidean 3D vector space}%
\end{array}%
$}}} \bigskip

\subsection{Time--Dependent Biodynamics}

Recall that in ordinary autonomous mechanics we have a \emph{configuration
manifold} $M$ (which denotes both $M_{rob}$ and $M_{hum}$), coordinated by $%
(x^{i})$, and the corresponding \textit{velocity phase--space manifold} is
its tangent bundle $TM$, coordinated by $(x^{i},\dot{x}^i)$. However, in
modern geometrical settings of non--autonomous mechanics, the configuration
manifold of time--dependent mechanics is a fibre bundle $\pi:M\rightarrow
\mathbb{R}$, called the \textit{configuration bundle}, coordinated by $%
(t,x^{i})$, where $t\in\mathbb{R}$ is a Cartesian coordinate on the time
axis $\mathbb{R}$ with the transition functions $t{^{\prime}}=t+$const. The
corresponding velocity phase--space is the 1--jet space $J^{1}(\mathbb{R},M)$%
, which admits the adapted coordinates $(t,x^i,x^i_t)=(t,x^i,\dot{x}^i)$.
Every \textit{dynamical equation} $\xi$ defines a \textit{connection} on the
affine jet bundle $J^{1}(\mathbb{R},M)\rightarrow M$, and vice versa \cite%
{GaneshADG}.

Given the configuration fibre bundle $M\rightarrow \mathbb{R}$ over the time
axis $\mathbb{R}$, we say that the \textit{$1-$jet manifold} $J^{1}(\mathbb{R%
},M)$ is defined as the set of equivalence classes $j_{t}^{1}s$ of sections $%
s^{i}:\mathbb{R}\rightarrow M$ of the bundle $M\rightarrow \mathbb{R}$,
which are identified by their values $s^{i}(t)$, as well as by the values of
their partial derivatives $\partial _{t}s^{i}=\partial _{t}s^{i}(t)$ at time
points $t\in \mathbb{R}$. The 1--jet manifold $J^{1}(\mathbb{R},M)$ is
coordinated by $(t,x^{i},\dot{x}^{i})$, so the 1--jets are local coordinate
maps
\begin{equation*}
j_{t}^{1}s:\mathbb{R}\rightarrow M,~t\mapsto (t,x^{i},\dot{x}^{i}).
\end{equation*}
Similarly, the \textit{$2-$jet manifold} $J^{2}(\mathbb{R},M)$ is the set of
equivalence classes $j_{t}^{2}s$ of sections $s^{i}:\mathbb{R}\rightarrow M$%
\ of the configuration bundle $\pi:M\rightarrow \mathbb{R}$, which are
identified by their values $s^{i}(t)$, as well as the values of their first
and second partial derivatives, $\partial _{t}s^{i}=\partial _{t}s^{i}(t)$
and $\partial _{tt}s^{i}=\partial _{tt}s^{i}(t)$, respectively, at time
points $t\in \mathbb{R}$. The 2--jet manifold $J^{2}(\mathbb{R},M)$ is
coordinated by $(t,x^{i},\dot{x}^{i},\ddot{x}^{i})$, so the 2--jets are
local coordinate maps
\begin{equation*}
j_{t}^{2}s:\mathbb{R}\rightarrow M,~t\mapsto (t,x^{i},\dot{x}^{i},\ddot{x}%
^{i}).
\end{equation*}

Given the configuration bundle $M\rightarrow \mathbb{R}$, coordinated by $%
(t,x^{i})$, and its 2--jet space $J^{2}(\mathbb{R},M)$, coordinated by $%
(t,x^i ,x_{t}^{i},x_{tt}^{i})$, any dynamical equation $\xi$ on the
configuration bundle $M\rightarrow \mathbb{R}$, which generalizes Lagrangian
equation (\ref{classic}),
\begin{equation}
\xi:\quad x_{tt}^{i}=\xi ^{i}(t,x^{i},x_{t}^{i})  \label{cqg5}
\end{equation}%
is equivalent to the \textit{geodesic equation} with respect to some affine
connection $\Gamma$ on the tangent bundle $TM\rightarrow M$,
\begin{equation*}
\dot{t}=1,\qquad \ddot{t}=0, \qquad \,\ddot{x}^{i}=\Gamma_{0}^{i}+%
\Gamma_{j}^{i}\dot{x}^{j},  \label{cqg11}
\end{equation*}%
which fulfills the conditions
\begin{equation}
\Gamma_{\alpha }^{0}=0,\qquad \xi
^{i}=\Gamma_{0}^{i}+x_{t}^{j}\Gamma_{j}^{i}\mid _{\dot{t}=1,\dot{x}%
^{i}=x_{t}^{i}}.  \label{cqg9}
\end{equation}

A holonomic connection $\xi$ is represented by the horizontal vector--field
on $J^{1}(\mathbb{R},M)$,
\begin{equation}
\xi =\partial _{t}+x_{t}^{i}\partial _{i}+\xi ^{i}(x^{\mu
},x_{t}^{i})\partial _{i}^{t}.  \label{a1.30}
\end{equation}

A dynamical equation $\xi $ is said to be \emph{conservative} if there
exists a trivialization $M\cong \mathbb{R}\times M$ such that the
vector--field $\xi $ (\ref{a1.30}) on $J^{1}(\mathbb{R},M)\cong \mathbb{R}%
\times TM$ is projectable onto $TM$. Then this projection
\begin{equation*}
\Xi _{\xi }=\dot{x}^{i}\partial _{i}+\xi ^{i}(x^{j},\dot{x}^{j})\dot{\partial%
}_{i}  \label{jp26}
\end{equation*}%
is a second--order dynamical equation on a typical fibre $M$ of $%
M\rightarrow \mathbb{R}$,
\begin{equation}
\ddot{x}^{i}=\Xi _{\xi }^{i}.  \label{nll1}
\end{equation}%
Conversely, every second--order dynamical equation $\Xi $ (\ref{nll1}) on a
manifold $M$ can be seen as a conservative dynamical equation
\begin{equation*}
\xi _{\Xi }=\partial _{t}+\dot{x}^{i}\partial _{i}+u^{i}\dot{\partial}_{i}
\label{jp27}
\end{equation*}%
on the trivial fibre bundle $\mathbb{R}\times M\rightarrow \mathbb{R}$.

\subsubsection{Nonautonomous Dissipative Hamiltonian Dynamics}

We can now formulate the time-dependent biomechanics \cite%
{tijLag,tijHam,tijRic} in which the biomechanical phase space is the
Legendre manifold\footnote{%
The maximum dimensional integral manifold of a certain diffeomorphism group
is called the Legendre manifold.} $\Pi $, endowed with the holonomic
coordinates $(t,y^{i},p_{i})$ with the transition functions
\begin{equation*}
p_{i}^{\prime }=\frac{\partial y^{j}}{\partial {y^{\prime }}^{i}}p_{j}.
\end{equation*}
$\Pi $ admits the canonical form ${\Lambda }$ given by
\begin{equation*}
{\Lambda }=dp_{i}\wedge dy^{i}\wedge dt\otimes \partial _{t}.
\end{equation*}
We say that a connection
\begin{equation*}
\gamma =dt\otimes (\partial _{t}+\gamma ^{i}\partial _{i}+\gamma
_{i}\partial ^{i})
\end{equation*}
on the bundle $\Pi \rightarrow X$ is \emph{locally Hamiltonian} if the
exterior form $\gamma \rfloor {\Lambda }$ is closed and Hamiltonian if the
form $\gamma \rfloor {\Lambda }$ is exact \cite{book}. A connection $\gamma $
is locally Hamiltonian iff it obeys the conditions:
\begin{equation*}
\partial ^{i}\gamma ^{j}-\partial ^{j}\gamma ^{i}=0,\quad \partial
_{i}\gamma _{j}-\partial _{j}\gamma _{i}=0,\quad \partial _{j}\gamma
^{i}+\partial ^{i}\gamma _{j}=0.
\end{equation*}

Note that every connection $\Gamma=dt\otimes(\partial_t +\Gamma^i\partial_i)$
on the bundle $Y\to X$ gives rise to the Hamiltonian connection $%
\widetilde\Gamma$ on $\Pi\to X$, given by
\begin{equation*}
\widetilde\Gamma =dt\otimes(\partial_t +\Gamma^i\partial_i
-\partial_j\Gamma^i p_i\partial^j).
\end{equation*}
The corresponding Hamiltonian form $H_\Gamma$ is given by
\begin{equation*}
H_\Gamma=p_idy^i -p_i\Gamma^idt.
\end{equation*}

Let $H$ be a \emph{dissipative Hamiltonian form} on $\Pi$, which reads:
\begin{equation}
H=p_idy^i-\mathcal{H} dt=p_idy^i -p_i\Gamma^idt -\widetilde{\mathcal{H}}%
_\Gamma dt.  \label{m46}
\end{equation}
We call $\mathcal{H}$ and $\widetilde{\mathcal{H}}$ in the decomposition (%
\ref{m46}) the \textit{Hamiltonian} and the \textit{Hamiltonian function}
respectively. Let $\gamma$ be a Hamiltonian connection on $\Pi\to X$
associated with the Hamiltonian form (\ref{m46}). It satisfies the relations
\cite{book}
\begin{eqnarray}
&&\gamma\rfloor{\Lambda} =dp_i\wedge dy^i+ \gamma_idy^i\wedge dt
-\gamma^idp_i\wedge dt = dH,  \notag \\
&&\gamma^i =\partial^i\mathcal{H}, \qquad \gamma_i=-\partial_i\mathcal{H}.
\label{m40}
\end{eqnarray}
From equations (\ref{m40}) we see that, in the case of biomechanics, one and
only one Hamiltonian connection is associated with a given Hamiltonian form.

Every connection $\gamma$ on $\Pi\to X$ yields the system of first--order
differential equations:
\begin{equation}
\dot{y}^i =\gamma^i, \qquad \dot{p}_i =\gamma_i.  \label{m170}
\end{equation}
They are called the \textit{evolution equations}. If $\gamma$ is a
Hamiltonian connection associated with the Hamiltonian form $H$ (\ref{m46}),
the evolution equations (\ref{m170}) become the \emph{dissipative
time-dependent Hamiltonian equations}:
\begin{eqnarray}
\dot{y}^i =\partial^i\mathcal{H}, \qquad \dot{p}_i =-\partial_i\mathcal{H}.
\label{m41}
\end{eqnarray}

In addition, given any scalar function $f$ on $\Pi$, we have the \textit{%
dissipative Hamiltonian evolution equation}
\begin{equation}
d_{H}f=(\partial_t +\partial^i\mathcal{H}\partial_i -\partial_i\mathcal{H}%
\partial^i)\,f,  \label{m59}
\end{equation}
relative to the Hamiltonian $\mathcal{H}$. On solutions $s$ of the
Hamiltonian equations (\ref{m41}), the evolution equation (\ref{m59}) is
equal to the total time derivative of the function $f$:
\begin{eqnarray*}
s^*d_{H}f=\frac{d}{dt}(f\circ s).
\end{eqnarray*}

\subsubsection{Neuro--Muscular Fitness Dynamics}

The dissipative Hamiltonian system (\ref{m41})--(\ref{m59}) is the basis for
our time\,\&\,fitness-dependent biomechanics. The scalar function $f$ in (%
\ref{m59}) on the biomechanical Legendre phase-space manifold $\Pi$ is now
interpreted as an \emph{individual neuro-muscular fitness function}. This
fitness function is a `determinant' for the performance of muscular drives
for the driven, dissipative Hamiltonian biomechanics. These muscular drives,
for all active DOF, are given by time\,\&\,fitness-dependent Pfaffian form: $%
F_i=F_i(t,y,p,f)$. In this way, we obtain our final model for
time\,\&\,fitness-dependent Hamiltonian biomechanics: {\large
\begin{eqnarray*}
\dot{y}^i &=& \partial^i\mathcal{H}, \\
\dot{p}_i &=& F_i-\partial_i\mathcal{H}, \\
d_{H}f &= &(\partial_t +\partial^i\mathcal{H}\partial_i -\partial_i\mathcal{H%
}\partial^i)\,f.
\end{eqnarray*}
}

Physiologically, the active muscular drives $F_i=F_i(t,y,p,f)$ consist of
\cite{GaneshSprSml}):\newline

\textbf{1. Synovial joint mechanics}, giving the first stabilizing effect to
the conservative skeleton dynamics, is described by the $(y,\dot{y})$--form
of the \textit{Rayleigh--Van der Pol's dissipation function}
\begin{equation*}
R=\frac{1}{2}\sum_{i=1}^{n}\,(\dot{y}^{i})^{2}\,[\alpha _{i}\,+\,\beta
_{i}(y^{i})^{2}],\quad
\end{equation*}
where $\alpha _{i}$ and $\beta _{i}$ denote dissipation parameters. Its
partial derivatives give rise to the viscous--damping torques and forces in
the joints
\begin{equation*}
\mathcal{F}_{i}^{joint}=\partial R/\partial \dot{y}^{i},
\end{equation*}
which are linear in $\dot{y}^{i}$ and quadratic in $y^{i}$.\newline

\textbf{2. Muscular mechanics}, giving the driving torques and forces $%
\mathcal{F}_{i}^{musc}=\mathcal{F}_{i}^{musc}(t,y,\dot{ y})$ with $%
(i=1,\dots ,n)$ for human biomechanics, describes the internal {excitation}
and {contraction} dynamics of \textit{equivalent muscular actuators} \cite%
{GaneshSprSml}.\newline

(a) The \textit{excitation dynamics} can be described by an impulse {%
force--time} relation
\begin{eqnarray*}
F_{i}^{imp} &=&F_{i}^{0}(1\,-\,e^{-t/\tau _{i}})\text{ \qquad if stimulation
}>0 \\
\quad F_{i}^{imp} &=&F_{i}^{0}e^{-t/\tau _{i}}\qquad \qquad \;\quad\text{if
stimulation }=0,\quad
\end{eqnarray*}
where $F_{i}^{0}$ denote the maximal isometric muscular torques and forces,
while $\tau _{i}$ denote the associated time characteristics of particular
muscular actuators. This relation represents a solution of the Wilkie's
muscular \textit{active--state element} equation \cite{Wilkie}
\begin{equation*}
\dot{\mu}\,+\,\Gamma \,\mu \,=\,\Gamma \,S\,A,\quad \mu (0)\,=\,0,\quad
0<S<1,
\end{equation*}
where $\mu =\mu (t)$ represents the active state of the muscle, $\Gamma $
denotes the element gain, $A$ corresponds to the maximum tension the element
can develop, and $S=S(r)$ is the `desired' active state as a function of the
motor unit stimulus rate $r$. This is the basis for biomechanical force
controller.\newline

(b) The \textit{contraction dynamics} has classically been described by
Hill's \textit{hyperbolic force--velocity} relation \cite{Hill}
\begin{equation*}
F_{i}^{Hill}\,=\,\frac{\left( F_{i}^{0}b_{i}\,-\,\delta _{ij}a_{i}\dot{y}%
^{j}\,\right) }{\left( \delta _{ij}\dot{y}^{j}\,+\,b_{i}\right) },\,\quad
\end{equation*}
where $a_{i}$ and $b_{i}$ denote {Hill's parameters}, corresponding to the
energy dissipated during the contraction and the phosphagenic energy
conversion rate, respectively, while $\delta _{ij}$ is the Kronecker's $%
\delta-$tensor.

In this way, human biomechanics describes the excitation/contraction
dynamics for the $i$th equivalent muscle--joint actuator, using the simple
impulse--hyperbolic product relation
\begin{equation*}
\mathcal{F}_{i}^{musc}(t,y,\dot{y})=\,F_{i}^{imp}\times F_{i}^{Hill}.\quad
\end{equation*}

\section{Hierarchical Control of Humanoid Robots}

\subsection{Spinal Control Level}

The \textit{force HBE servo--controller} is formulated as affine control
Hamiltonian--systems (\ref{af1}--\ref{af2}) (with possible extensions along
the lines of the previous section), which resemble an {autogenetic motor
servo} \cite{Houk}, acting on the spinal--reflex level of the human
locomotion control. A voluntary contraction force $F$ of human skeletal
muscle is reflexly excited (positive feedback $+F^{-1}$) by the responses of
its {spindle receptors} to stretch and is reflexly inhibited (negative
feedback $-F^{-1}$) by the responses of its {Golgi tendon organs} to
contraction. Stretch and unloading reflexes are mediated by combined actions
of several autogenetic neural pathways, forming the so--called $\mathbf{`}$%
motor servo.' The term $\mathbf{`}$autogenetic' means that the stimulus
excites receptors located in the same muscle that is the target of the
reflex response. The most important of these muscle receptors are the
primary and secondary endings in the muscle--spindles, which are sensitive
to length change -- positive length feedback $+F^{-1}$, and the Golgi tendon
organs, which are sensitive to contractile force -- negative force feedback $%
-F^{-1}$.

The gain $G$ of the length feedback $+F^{-1}$ can be expressed as the {%
positional stiffness} (the ratio $G\approx S=dF/dx$ of the force--$F$ change
to the length--$x$ change) of the muscle system. The greater the stiffness $%
S $, the less the muscle will be disturbed by a change in load. The
autogenetic circuits $+F^{-1}$ and $-F^{-1}$ appear to function as {%
servoregulatory loops} that convey continuously graded amounts of excitation
and inhibition to the large ({alpha}) skeletomotor neurons. Small ({gamma})
fusimotor neurons innervate the contractile poles of muscle spindles and
function to modulate spindle--receptor discharge.

\subsection{Cerebellum--Like Velocity and Jerk Control}

{Nonlinear velocity} and {jerk} (time derivative of acceleration) {%
servo--controllers} \cite{GaneshSprSml}, developed using the Lie--derivative
formalism \cite{SIAM}, resemble self--stabilizing and adaptive tracking
action of the cerebellum \cite{HoukBarto}. By introducing the vector--fields
$f$ and $g$, given respectively by
\begin{equation*}
f=\left( \partial _{p_{i}}H_{0},\,-\partial _{q^{i}}H_{0}\right) ,\qquad
g=\left( -\partial _{p_{i}}H^{j},\,\partial _{q^{i}}H^{j}\right) ,
\end{equation*}%
we obtain the affine controller in the standard nonlinear MIMO--system form
(see \cite{GaneshADG})
\begin{equation}
\dot{x}_{i}=f(x)+g(x)\,u_{j}.  \label{MIMO}
\end{equation}

Finally, using the {Lie derivative formalism} \cite{GaneshADG}\footnote{%
Let $F(M)$ denote the set of all smooth (i.e., $C^{\infty }$) real valued
functions $f:M\rightarrow \mathbb{R}$ on a smooth manifold $M$, $V(M)$ --
the set of all smooth vector--fields on $M$, and $V^{\ast }(M)$ -- the set
of all differential one--forms on $M$. Also, let the vector--field $\zeta
\in V(M)$ be given with its local flow $\phi _{t}:M\rightarrow M$ such that
at a point $x\in M$, $\frac{d}{dt}|_{t=0}\,\phi _{t}x=\zeta (x)$, and $\phi
_{t}^{\ast }$ representing the {pull--back} by $\phi _{t}$. The {Lie
derivative} differential operator $\mathcal{L}_{\zeta }$ is defined:
\par
(i) on a function $f\in F(M)$ as
\begin{equation*}
\mathcal{L}_{\zeta }:F(M)\rightarrow F(M),\qquad \mathcal{L}_{\zeta }f=\frac{%
d}{dt}(\phi _{t}^{\ast }f)|_{t=0},
\end{equation*}%
\par
(ii) on a vector--field $\eta \in V(M)$ as
\begin{equation*}
\mathcal{L}_{\zeta }:V(M)\rightarrow V(M),\qquad \mathcal{L}_{\zeta }\eta =%
\frac{d}{dt}(\phi _{t}^{\ast }\eta )|_{t=0}\equiv \lbrack \zeta ,\eta ]
\end{equation*}%
-- the {Lie bracket}, and
\par
(iii) on a one--form $\alpha \in V^{\ast }(M)$ as
\begin{equation*}
\mathcal{L}_{\zeta }:V^{\ast }(M)\rightarrow V^{\ast }(M),\qquad \mathcal{L}%
_{\zeta }\alpha =\frac{d}{dt}(\phi _{t}^{\ast }\alpha )|_{t=0}.
\end{equation*}%
In general, for any smooth tensor field $\mathbf{T}$ on $M$, the Lie
derivative $\mathcal{L}_{\zeta }\mathbf{T}$ geometrically represents a
directional derivative of $\mathbf{T}$ along the flow $\phi_{t}$.} and
applying the {constant relative degree} $r$ to all HB joints, the {control
law} for asymptotic tracking of the reference outputs $%
o_{R}^{j}=o_{R}^{j}(t) $ could be formulated as (generalized from \cite%
{Isidori})
\begin{equation}
u_{j}=\frac{\dot{o}_{R}^{(r)j}-L_{f}^{(r)}H^{j}+%
\sum_{s=1}^{r}c_{s-1}(o_{R}^{(s-1)j}-L_{f}^{(s-1)}H^{j})}{%
L_{g}L_{f}^{(r-1)}H^{j}},  \label{CtrLaw}
\end{equation}
where $c_{s-1}$ are the coefficients of the linear differential equation of
order $r$ for the {error function} $e(t)=x^{j}(t)-o_{R}^{j}(t)$
\begin{equation*}
e^{(r)}+c_{r-1}e^{(r-1)}+\dots+c_{1}e^{(1)}+c_{0}e=0.
\end{equation*}

The control law (\ref{CtrLaw}) can be implemented symbolically in $%
Mathematica^{TM}$ in the following three steps:

1. Symbolic functions defining the gradient and Lie derivatives:\footnote{%
This is the code in $Mathematica^{TM}$ version 7.}
\begin{eqnarray*}
\text{Grad}[s\_,x\_\text{List}] &:&=(D[s,\#1]\&)/@x; \\
\text{LieDer}[v\_\text{List},s\_,x\_\text{List}] &:&=\text{Grad}[s,x]\cdot v;
\\
\text{KLieDer}[v\_\text{List},s\_,x\_\text{List},k\_] &:&= \\
\text{Block}[\{t\},p &:&=s;~\text{If}[k==0,p=s,\text{Do}[p=\text{LieDer}%
[v,p,x],\{k\}]];p];
\end{eqnarray*}

2. Control law defined (for simplicity, we show here only the first--order
control law):%
\begin{eqnarray*}
u[t\_] =(-\text{LieDer}[F,y,X]+D[yR[t],t]+ \alpha (yR[t]-y))/\text{LieDer}%
[g,y,X];
\end{eqnarray*}

3. Example for the reference output $yR[t],$ with the final time Tfin:
\begin{eqnarray*}
yR[t\_] =\text{If}[t<=\text{Tfin}/2,5(1-\text{e}^{-5t}), (5(1-\text{e}%
^{-5t}))/\text{e}^{(5(t-\text{Tfin}/2)})];
\end{eqnarray*}

The affine nonlinear MIMO control system (\ref{MIMO}) with the
Lie--derivative control law (\ref{CtrLaw}) resembles the self--stabilizing
and synergistic output tracking action of the human cerebellum \cite{NeuFuz}%
. To make it adaptive (and thus more realistic), instead of the `rigid'
controller (\ref{CtrLaw}), we can use the {adaptive Lie--derivative
controller}, as explained in the seminal paper on geometrical nonlinear
control \cite{SI}.

\subsection{Cortical--Like Fuzzy--Topological Control}

For the purpose of our cortical control, the dominant, rotational part of
the human configuration manifold $M^{N}$, could be first, reduced to an $N$--%
{torus}, and second, transformed to an $N$--{cube} (`hyper--joystick'),
using the following topological techniques (see \cite{GaneshADG}).\footnote{%
This top control level has not yet been implemented. The main reason for
this is its high dimensionality. For example, the Human Biodynamics Engine
simulator has 270 degrees of freedom (both rotational and translational).
Its rotational part includes 135 individual Lie-derivative controllers. The
integration of so many individual controllers is a nontrivial problem that
is currently beyond the capacity of pure fuzzy control.}

Let $S^{1}$ denote the constrained unit circle in the complex plane, which
is an Abelian Lie group. Firstly, we propose two reduction homeomorphisms,
using the Cartesian product of the constrained $SO(2)-$groups:
\begin{equation*}
SO(3)\approx SO(2)\times SO(2)\times SO(2)\qquad\text{and} \qquad
SO(2)\approx S^{1}.
\end{equation*}

Next, let $I^{N}$ be the unit cube $[0,1]^{N}$ in $\mathbb{R}^{N}$ and `$%
\sim $' an equivalence relation on $\mathbb{R}^{N}$ obtained by `gluing'
together the opposite sides of $I^{N}$, preserving their orientation.
Therefore, $M^{N}$ can be represented as the quotient space of $\mathbb{R}%
^{N}$ by the space of the integral lattice points in $\mathbb{R}^{N}$, that
is an oriented and constrained $N$--dimensional torus $T^{N}$:
\begin{equation}
{\mathbb{R}^{N}/{Z}^{N}}\approx \,\prod_{i=1}^{N}S_{i}^{1}\equiv
\{(q^{i},\,i=1,\dots ,N):\mbox{mod}2\pi \}=T^{N}.  \label{torus}
\end{equation}%
Its {Euler--Poincar\'{e} characteristic} is (by the {De Rham theorem}) both
for the configuration manifold $T^{N}$ and its {momentum phase--space} $%
T^{\ast }T^{N}$ given by (see \cite{GaneshADG})
\begin{equation*}
\chi (T^{N},T^{\ast }T^{N})=\sum_{p=1}^{N}(-1)^{p}b_{p}\,,
\end{equation*}%
where $b_{p}$ are the {Betti numbers} defined as
\begin{align*}
b^{0}& =1,\, \\
b^{1}& =N,\dots b^{p}={\binom{N}{p}},\dots b^{N-1}=N, \\
b^{N}& =1,\qquad \qquad (0\leq p\leq N).
\end{align*}

Conversely by `ungluing' the configuration space we obtain the primary unit
cube. Let `$\sim^{\ast}$' denote an equivalent decomposition or `ungluing'
relation. According to Tychonoff's {product--topology theorem} \cite%
{GaneshADG}, for every such quotient space there exists a `selector' such
that their quotient models are homeomorphic, that is, $T^{N}/\sim^{\ast}%
\approx A^{N}/\sim^{\ast}$. Therefore $I_{q}^{N}$ represents a `selector'
for the configuration torus $T^{N}$ and can be used as an $N$--directional `$%
\hat{q}$--command--space' for the {feedback control} (FC). Any subset of
degrees of freedom on the configuration torus $T^{N}$ representing the
joints included in HB has its simple, rectangular image in the rectified $%
\hat{q}$--command space -- selector $I_{q}^{N}$, and any joint angle $q^{i}$
has its rectified image $\hat{q}^{i}$.

In the case of an end--effector, $\hat{q}^{i}$ reduces to the position
vector in external--Cartesian coordinates $z^{r}\,(r=1,\dots ,3)$. If
orientation of the end--effector can be neglected, this gives a topological
solution to the standard inverse kinematics problem.

Analogously, all momenta $\hat{p}_{i}$ have their images as rectified
momenta $\hat{p}_{i}$ in the $\hat{p}$--command space -- selector $I_{p}^{N}$%
. Therefore, the total momentum phase--space manifold $T^{\ast }T^{N}$
obtains its `cortical image' as the $\widehat{(q,p)}$--command space, a
trivial $2N$--dimensional bundle $I_{q}^{N}\times I_{p}^{N}$.

Now, the simplest way to perform the feedback FC on the cortical $\widehat{%
(q,p)}$--command space $I_{q}^{N}\times I_{p}^{N}$, and also to mimic the
cortical--like behavior, is to use the $2N$-- dimensional fuzzy--logic
controller, in much the same way as in the popular `inverted pendulum'
examples (see \cite{Kosko}).

We propose the fuzzy feedback--control map $\Xi $ that maps all the
rectified joint angles and momenta into the feedback--control one--forms
\begin{equation}
\Xi :(\hat{q}^{i}(t),\,\hat{p}_{i}(t))\mapsto u_{i}(t,q,p),  \label{map}
\end{equation}%
so that their corresponding universes of discourse, $\hat{Q}^{i}=(\hat{q}%
_{max}^{i}-\hat{q}_{min}^{i})$, $\hat{P}_{i}=(\hat{p}_{i}^{max}-\hat{p}%
_{i}^{min})$ and $\hat{U}{}_{i}=(u_{i}^{max}-u_{i}^{min})$, respectively,
are mapped as
\begin{equation}
\Xi :\prod_{i=1}^{N}\hat{Q}^{i}\times \prod_{i=1}^{N}\hat{P}_{i}\rightarrow
\prod_{i=1}^{N}{}\hat{U}{}_{i}.  \label{map1}
\end{equation}
\begin{figure}[htb]
\centering \includegraphics[width=10cm]{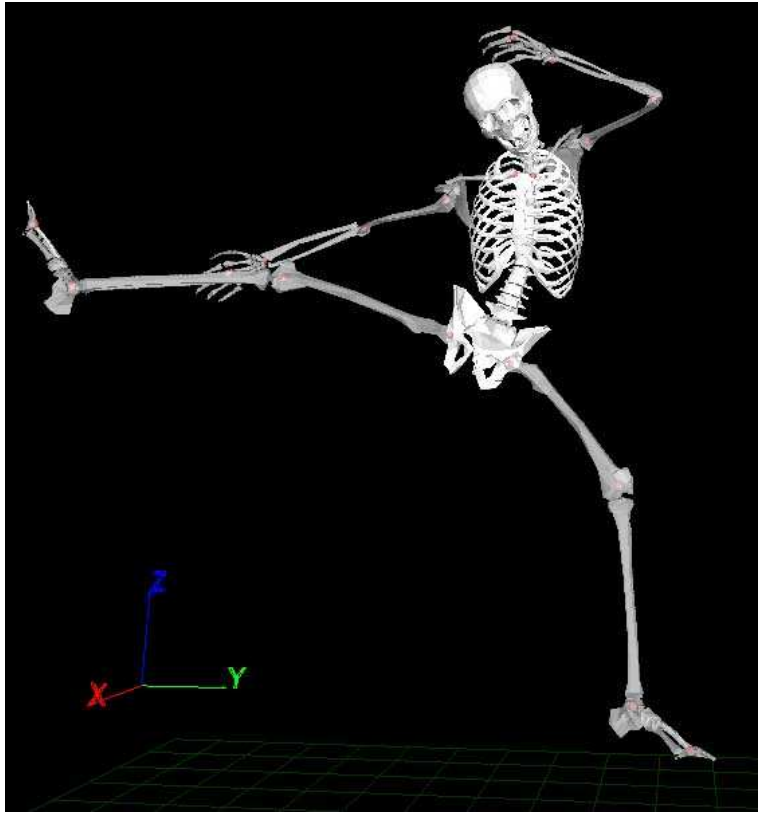}
\caption{The HBE simulating a jump-kick: a 3D viewer.}
\label{JumpKick}
\end{figure}

The $2N$--dimensional map $\Xi$ (\ref{map},\ref{map1}) represents a {fuzzy
inference system}, defined by \cite{NeuFuz}:

\begin{enumerate}
\item {Fuzzification} of the crisp {rectified} and {discretized} angles,
momenta and controls using Gaussian--bell membership functions
\begin{equation*}
\mu _{k}(\chi )=\exp[-\frac{(\chi -m_{k})^{2}}{2\sigma _{k}}],\qquad
(k=1,2,\dots ,9),
\end{equation*}%
where $\chi \in D$ is the common symbol for $\hat{q}^{i}$, $\hat{p}_{i}$ and
$u_{i}(q,p)$ and $D$ is the common symbol for $\hat{Q}^{i},\hat{P}_{i}$ and $%
{}_{i}$; the mean values $m_{k}$ of the nine partitions of each universe of
discourse $D$ are defined as $m_{k}=\lambda _{k}D+\chi _{min}$, with
partition coefficients $\lambda _{k}$ uniformly spanning the range of $D$,
corresponding to the set of nine linguistic variables $L=%
\{NL,NB,NM,NS,ZE,PS,PM$, $PB,PL\}$; standard deviations are kept constant $%
\sigma _{k}=D/9$. Using the linguistic vector $L$, the $9\times 9$ FAM
(fuzzy associative memory) matrix (a `linguistic phase--plane'), is
heuristically defined for each human joint, in a symmetrical weighted form
\begin{equation*}
\mu _{kl}=\varpi _{kl}\,exp\{-50[\lambda _{k}+u(q,p)]^{2}\},\qquad
(k,l=1,...,9)
\end{equation*}%
with weights $\varpi _{kl}\in \{0.6,0.6,0.7,0.7,0.8,0.8,0.9,0.9,1.0\}$.

\item {Mamdani inference} is used on each FAM--matrix $\mu_{kl}$ for all
human joints:\newline
(i) $\mu(\hat{q}^{i})$ and $\mu(\hat{p}_{i})$ are combined inside the fuzzy
IF--THEN rules using AND (Intersection, or Minimum) operator,
\begin{equation*}
\mu_{k}[\bar{u}_{i}(q,p)]=\min_{l}\{\mu_{kl}(\hat{q}^{i}),\,\mu_{kl}(\hat {p}%
_{i})\}.
\end{equation*}
(ii) the output sets from different IF--THEN rules are then combined using
OR (Union, or Maximum) operator, to get the final output, fuzzy--covariant
torques,
\begin{equation*}
\mu[u_{i}(q,p)]=\max_{k}\{\mu_{k}[\bar{u}_{i}(q,p)]\}.
\end{equation*}

\item {Defuzzification} of the fuzzy controls $\mu \lbrack u_{i}(q,p)] $
with the `center of gravity' method
\begin{equation*}
u_{i}(q,p)=\frac{\int \mu \lbrack u_{i}(q,p)]\,du_{i}}{\int du_{i}},
\end{equation*}%
to update the crisp feedback--control one--forms $u_{i}=u_{i}(t,q,p)$.
\end{enumerate}

\begin{figure}[htb]
\centering \includegraphics[width=12cm]{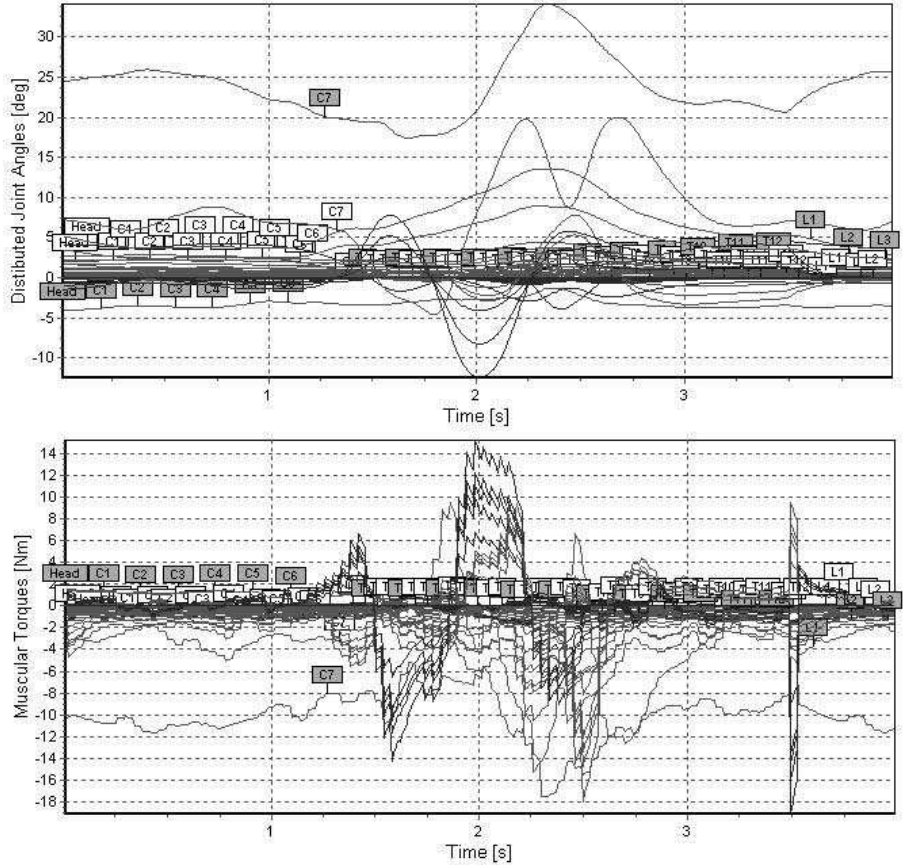}
\caption{The HBE simulating a jump-kick: calculating joint angles and
muscular torques.}
\label{AnglesTorques}
\end{figure}
\begin{figure}[htb]
\centering \includegraphics[width=8cm]{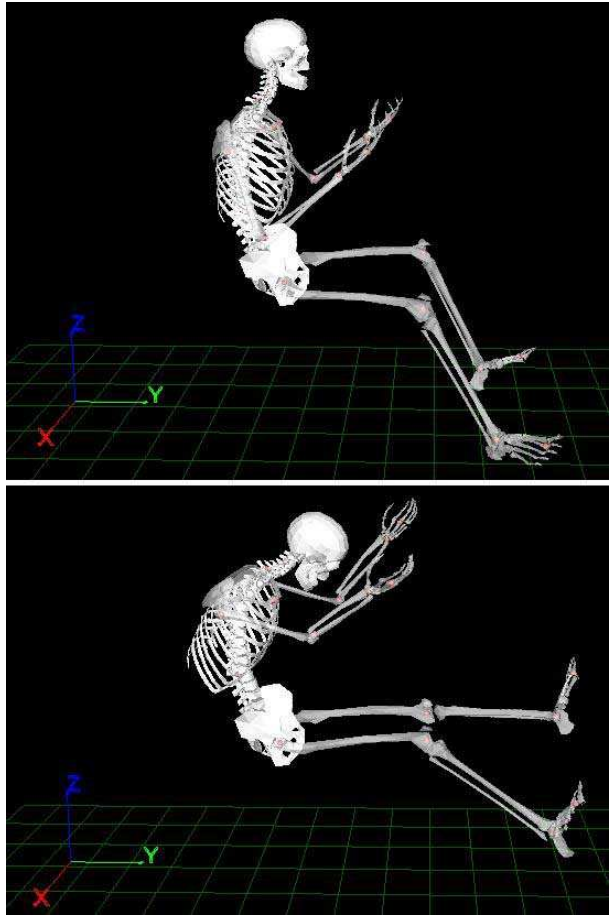}
\caption{The HBE simulating the frontal road-vehicle crash into the fixed
wall with a speed of 70 km/h: before the impact (up) and 0.12 s after the
impact.}
\label{RVcrash}
\end{figure}

Now, it is easy to make this top--level controller {adaptive}, simply by {%
weighting} both the above fuzzy--rules and membership functions, by the use
of any standard competitive neural--network (see, e.g., \cite{Kosko}).
Operationally, the construction of the cortical $\widehat{(q,p)}$--command
space $I_{q}^{N}\times I_{p}^{N}$ and the $2N$--dimensional feedback map $%
\Xi $ (\ref{map},\ref{map1}), mimic the regulation of the {motor conditioned
reflexes} by the motor cortex \cite{HoukBarto}.

It has been implicitly assumed that close resemblance of hierarchical
control structures presented in this section with the corresponding human
neuro-physiological control mechanisms would assure the necessary overall
stability of biodynamics. However, in future work, these control structures
need to be properly analyzed, starting with Lyapunov stability criteria.

\section{Simulation Examples}

In this section we give several simulation examples of the sophisticated
virtual humanoid called Human Biodynamics Engine (HBE), designed at Defence
Science \& Technology Organisation, Australia. The HBE is a sophisticated
human neuro-musculo-skeletal dynamics simulator, based on generalized
Lagrangian and Hamiltonian mechanics and Lie-derivative nonlinear control.
It includes 270 active degrees of freedom (DOF), while fingers are not
separately modelled: 135 rotational DOF are considered active, and 135
translational DOF are considered passive. The HBE incorporates both forward
and inverse dynamics, as well as two neural--like control levels. Active
rotational joint dynamics is driven by 270 nonlinear muscular actuators,
each with its own excitation--contraction dynamics (following traditional
Hill--Hatze biomechanical models). Passive translational joint dynamics
models visco-elastic properties of inter-vertebral discs, joint tendons and
muscular ligaments as a nonlinear spring-damper system. The lower neural
control level resembles spinal--reflex positive and negative force
feedbacks, resembling stretch and Golgi reflexes, respectively. The higher
neural control level mimics cerebellum postural stabilization and velocity
target-tracking control. The HBE's core is the full spine simulator,
considering human spine as a chain of 26 flexibly--coupled rigid bodies
(formally, the product of 26 SE(3)--groups). The HBE includes over 3000 body
parameters, all derived from individual user data, using standard
biomechanical tables. The HBE incorporates a new theory of soft
neuro-musculo-skeletal injuries, based on the concept of the local
rotational and translational {jolts}, which are the time rates of change of
the total forces and torques localized in each joint at a particular time
instant.
\begin{figure}[htb]
\centering \includegraphics[width=12cm]{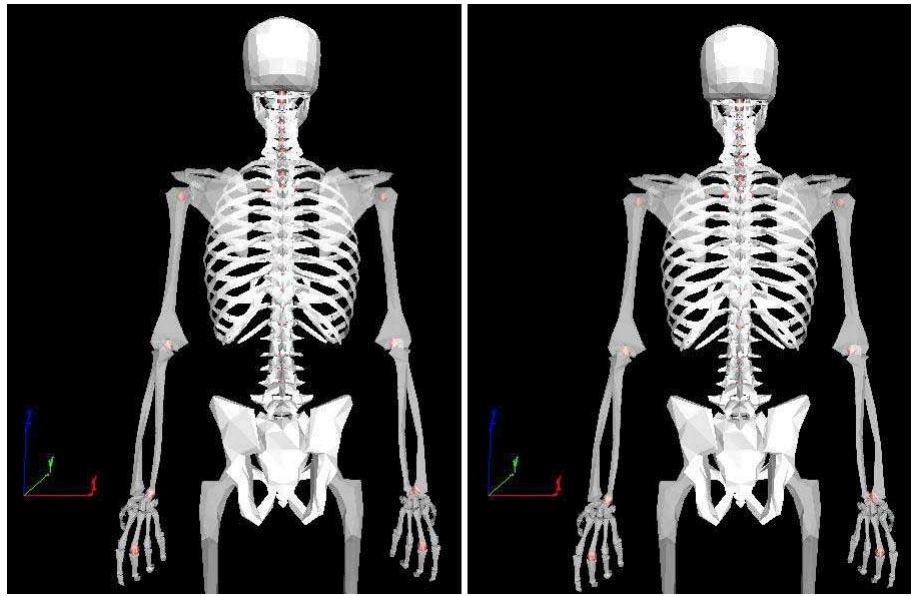}
\caption{The HBE simulating an effect of an aircraft pilot-seat ejection to
human spine compression: before the seat ejection (left) and after ejection
(right).}
\label{Ejection}
\end{figure}
\begin{figure}[htb]
\centering \includegraphics[width=12cm]{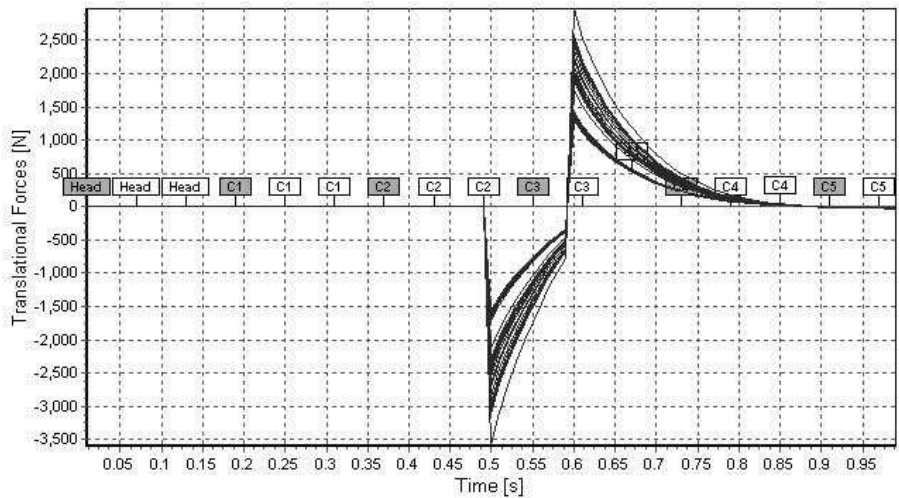}
\caption{The HBE calculating translational forces distributed along the
spinal joints during the seat ejection.}
\label{EjectForces}
\end{figure}

The first version of the HBE simulator had the full human-like skeleton,
driven by the generalized Hamiltonian dynamics (including muscular
force-velocity and force-time curves) and two levels of reflex-like motor
control (simulated using the Lie derivative formalism) \cite%
{SIAM,GaneshSprSml}. It had 135 purely rotational DOF, strictly following
Figure \ref{SpineSO(3)}. It was created for prediction and prevention of
musculo-skeletal injuries occurring in the joints, mostly spinal
(intervertebral). Its performance looked kinematically realistic, while it
was not possible to validate the driving torques. It included a small
library of target movements which were followed by the HBE's Lie--derivative
controllers with efficiency of about 90\% (see Figures \ref{JumpKick} and %
\ref{AnglesTorques}).
\begin{figure}[htb]
\centering \includegraphics[width=14cm]{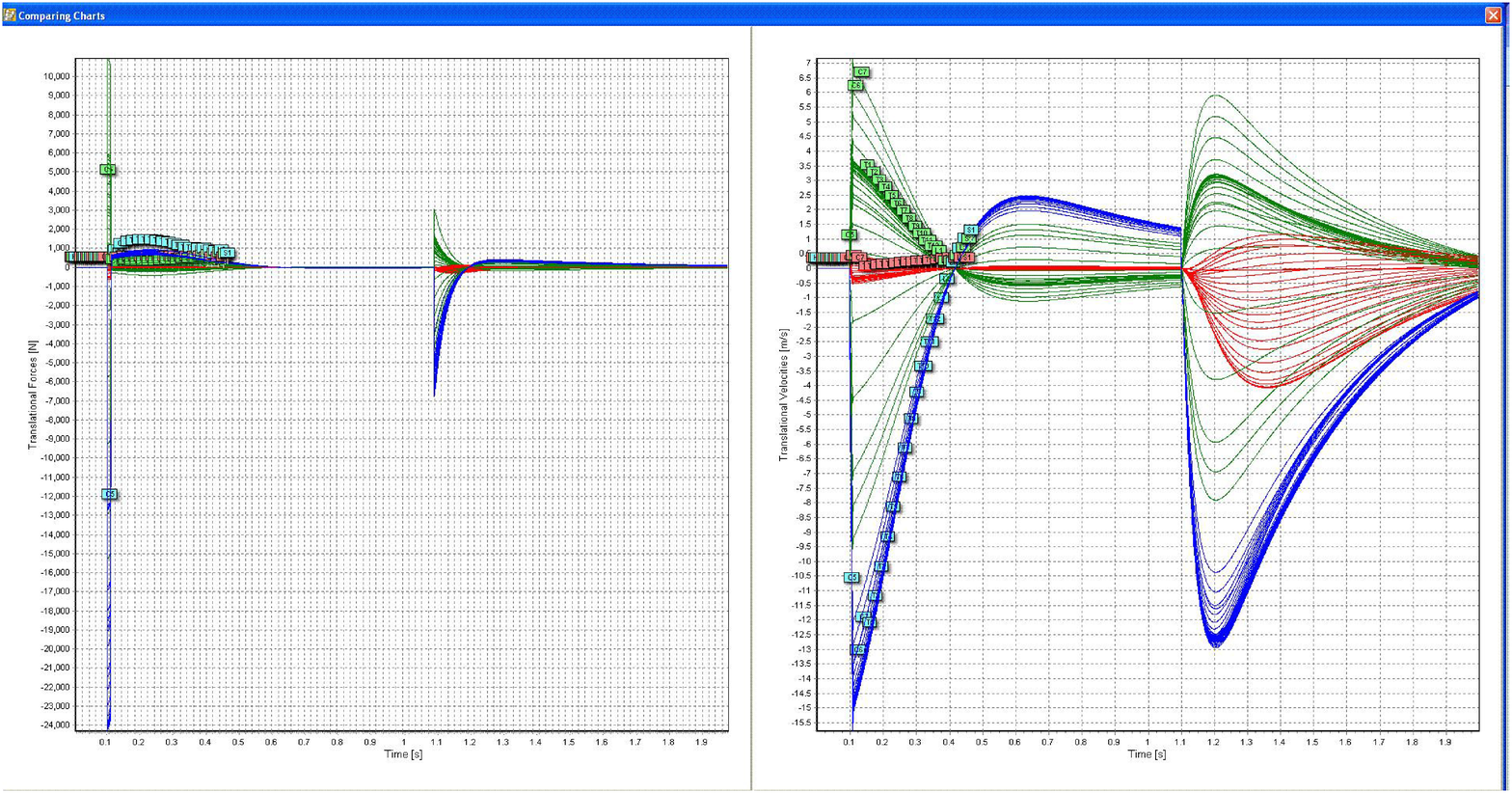}
\caption{The HBE simulating a double-impact effect of a land-mine blast
under an armor-protected vehicle to a hypothetical passenger's body.}
\label{2charts}
\end{figure}

The HBE also includes a generic crash simulator, based on the simplified
road-vehicle impact simulator (see Figure \ref{RVcrash}). While implementing
the generic crash simulator, it became clear that purely rotational joint
dynamics would not be sufficient for the realistic prediction of
musculo-skeletal injuries. In particular, to simulate the action of a
Russian aircraft ejection-seat currently used by the American space shuttle,
we needed to implement micro translations in the intervertebral joints (see
Figures \ref{Ejection} and \ref{EjectForces}). This is because the seat
provides full body restraint and hence the ejection rockets firing with 15 g
per .15 s cause pure compression of the spine (without any bending).

\begin{figure}[h]
\centerline{\includegraphics[width=8cm]{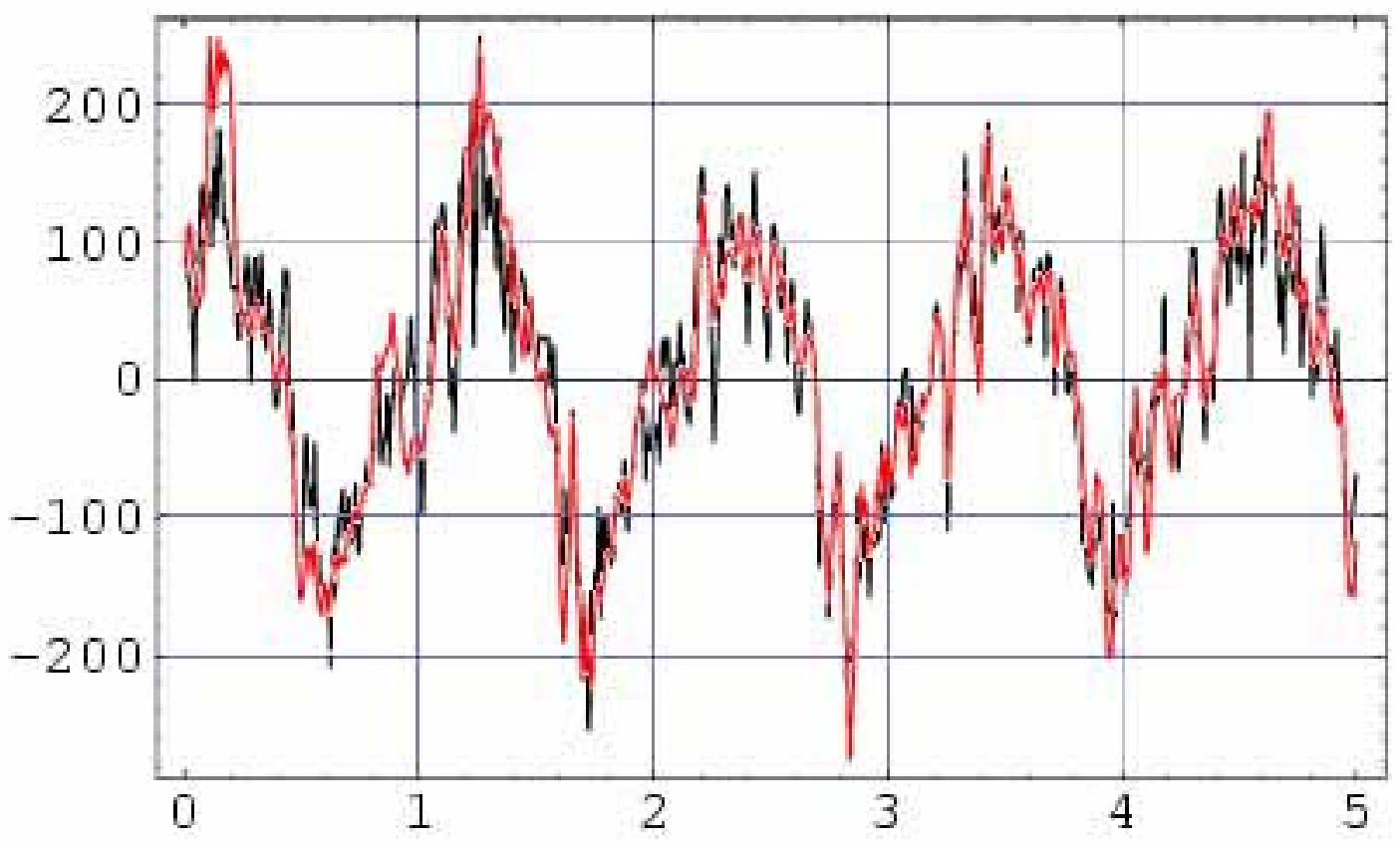}}
\caption{{\protect\small Matching the `Vicon' output with the `HBE' output
for the right-hip angular velocity around the dominant X-axis, while walking
on a treadmill with a speed of 4 km/h.}}
\label{WalkingValid}
\end{figure}

Finally, the HBE includes the defence-specific land-mine crash simulator. It
is calibrated on a hypothetical double-impact under the armor-protected
military vehicle, including:

\begin{enumerate}
\item A land-mine blast of 350g with a duration of 5ms;

\item A 1s pause when the hypothetical vehicle is in the air; ~and

\item The vehicle hard landing with an acceleration of 100g and a duration
1s.
\end{enumerate}

The HBE calculates full rotational and translational dynamics caused by the
land-mine double-impact in extreme force/time scales (including both linear
and angular displacements, velocities, forces and jolts in all human joints
(see Figure \ref{2charts})). The variations of the applied g-forces and
durations of the two impacts can be simulated, to see the differences in
their effects on the hypothetical passenger's body.

In this way a full rotational + translational biodynamics simulator has been
created with 270 DOF in total (not representing separate fingers). The
`HBE-simulator' has been kinematically validated (see Figure \ref%
{WalkingValid}) against the standard biomechanical gait-analysis system
`Vicon' \cite{Robertson}.

\end{document}